\documentclass[11pt]{article}   	
\usepackage{geometry,amsmath}  
\usepackage{graphicx}				
\usepackage{booktabs}
\usepackage{amssymb,tensor,stmaryrd,mathrsfs,bm,slashed,xcolor,cite,authblk}
\usepackage[colorlinks, linkcolor=blue, anchorcolor=blue, citecolor=blue ]{hyperref}
\allowdisplaybreaks[3]
\newcommand{\nn}{\nonumber}

\newcommand{\p}{\partial}
\newcommand{\tr}{\text{tr}}
\newcommand{\td}{\text{d}}
\newcommand{\LCGamma}{\mathring{\Gamma}}

\newcommand{\LCnabla}{\mathring{\nabla}}

\newcommand{\un}{\underline}
\newcommand{\E}{\text{e}}

\title{\huge Obstruction Tensors in Weyl Geometry\\and Holographic Weyl Anomaly}
\author{Weizhen Jia\footnote{weizhen2@illinois.edu}\,\, and Manthos Karydas\footnote{karydas2@illinois.edu}}
\affil{\small Department of Physics, University of Illinois,
 	1110 West Green St., Urbana IL 61801, U.S.A.}
\date{}
\begin{document}
\maketitle

\begin{abstract}
Recently a generalization of the Fefferman-Graham gauge for asymptotically locally AdS spacetimes, called the Weyl-Fefferman-Graham (WFG) gauge, has been proposed. It was shown that the WFG gauge induces a Weyl geometry on the conformal boundary. The Weyl geometry consists of a metric and a Weyl connection. Thus, this is a useful setting for studying dual field theories with background Weyl symmetry. Working in the WFG formalism, we find the generalization of obstruction tensors, which are Weyl-covariant tensors that appear as poles in the Fefferman-Graham expansion of the bulk metric for even boundary dimensions. We see that these Weyl-obstruction tensors can be used as building blocks for the Weyl anomaly of the dual field theory. We then compute the Weyl anomaly for $6d$ and $8d$ field theories in the Weyl-Fefferman-Graham formalism, and find that the contribution from the Weyl structure in the bulk appears as cohomologically trivial modifications. Expressed in terms of the Weyl-Schouten tensor and extended Weyl-obstruction tensors, the results of the holographic Weyl anomaly up to $8d$ also reveal hints on its expression in any dimension. 
\end{abstract}

\newpage
\begingroup
\hypersetup{linkcolor=black}
\tableofcontents
\endgroup

\newpage
\section{Introduction}
\label{Intro}
There is an important fact about the asymptotic AdS geometry: the conformal boundary of a $(d+1)$-dimensional asymptotically locally AdS (AlAdS) spacetime carries not a metric but a conformal class of metrics, i.e.\ the boundary enjoys Weyl symmetry. This is due to the fact that the asymptotic boundary is formally located at conformal infinity\cite{Penrose:1962ij}. In holographic theories\cite{Maldacena:1997re}, the  (background) Weyl symmetry is implied by  diffeomorphism invariance in the bulk spacetime (called \emph{Weyl diffeomorphism}). Usually when discussing AdS/CFT, one picks a specific representative of the conformal class. For example, the most commonly used choice for studying the conformal boundary of an AlAdS spacetime is the Fefferman-Graham (FG) gauge \cite{Feffe,Fefferman2011}. However, the FG  gauge explicitly breaks the Weyl symmetry by fixing a specific boundary metric.
\par
In a suitable coordinate system $\{z,x^\mu\}$ ($\mu=0,\cdots,d-1$), the metric of any $(d+1)$-dimensional AlAdS spacetime can be expanded with respect to the bulk coordinate $z$ into two series, called the Fefferman-Graham expansion\cite{Ciambelli:2019bzz,Leigh}. The first series has the boundary metric in the leading order, while the subleading terms are determined by the bulk equations of motion; the leading order of the second series represents the vacuum expectation value of the energy-momentum tensor operator of the boundary field theory, which cannot be determined in the absence of an interior boundary condition\cite{Leigh}.
\par
When the spacetime dimension is odd, both series in the FG expansion are power series to infinite order; however, in an even-dimensional spacetime, a logarithmic term will occur at order $O(z^{d-2})$, causing an obstruction to the power series expansion\cite{graham2005ambient}. This logarithmic term in $d=2k$ (with $k$ an integer) gives rise to the \emph{obstruction tensor} ${\cal O}^{(2k)}_{\mu\nu}$. The obstruction tensor was first proposed in \cite{Feffe} as a symmetric traceless tensor of type $(0,2)$, which is Weyl-covariant with Weyl weight $2k-2$ ($k\geqslant2$), and was precisely defined using the ambient metric in \cite{graham2005ambient} (see also \cite{Fefferman2011}). It is also convenient to define the \emph{extended obstruction tensor}\cite{GRAHAM20091956} which has a pole at $d=2k$, and whose residue gives rise to the obstruction tensor. The obstruction tensor for $d=4$ is also known as the Bach tensor\cite{Bach}, which is the only Weyl-covariant tensor in $4d$ that is algebraically independent of the Weyl tensor. It has been shown in \cite{graham2005ambient} that the only irreducible Weyl-covariant tensors in $2k$-dimension with $k\geqslant2$ are the obstruction tensor ${\cal O}^{(2k)}_{\mu\nu}$ and the Weyl tensor (which has weight $0$), while in any odd dimension $d=2k+1$ with $k\geqslant2$ the Weyl tensor is the only one (in $3d$ where the Weyl tensor becomes trivial, it is the Cotton tensor). 
\par
The origin of the obstruction tensor in the FG expansion is that the two series will mix if the spacetime dimension $d$ is even, and the solution to the equations of motion encounters a pole. Hence, another way to formulate the FG expansion is to use the technique of dimensional regularization, i.e.\ to regard $d$ as a variable (formally complex)\cite{Ciambelli:2019bzz,Leigh}. Using this formulation, in this paper we will describe a practical way of reading off the obstruction tensor from the pole of the FG expansion in an even dimension.
\par
Even though the FG gauge is quite convenient to use, the Weyl symmetry in the boundary will be broken when the boundary metric is fixed. More specifically, one can introduce a Penrose-Brown-Henneaux (PBH) transformation \cite{Imbimbo_2000,Bautier:2000mz,Rooman:2000ei} in the bulk and induce a Weyl transformation on the boundary, but the subleading terms in the $z$-expansion will not transform in a Weyl-covariant way if the form of the FG ansatz is to be preserved. In order to resolve this issue, one can relax the FG ansatz of the bulk metric to the Weyl-Fefferman-Graham (WFG) ansatz \cite{Ciambelli:2019bzz}. In the WFG gauge, the form of the bulk metric is preserved under a Weyl diffeomorphism, and all the terms in the $z$-expansion transform in a Weyl-covariant way, which brings a powerful reorganization of the holographic dictionary. Unlike the FG gauge, where the bulk Levi-Civita (LC) connection induces on the conformal boundary also a LC connection (of the boundary metric), in the WFG gauge, the bulk LC connection gives a Weyl connection on the boundary\cite{Ciambelli:2019bzz}. Having the induced metric together with the Weyl connection, the bulk geometry induces on the boundary a Weyl-covariant geometry\cite{Folland:1970,Hall:1992,Scholz2018}. 
\par
On the boundary, the induced metric and the Weyl connection act as non-dynamical backgrounds of the dual quantum field theory. Similar to the FG case, the metric is the source of the energy-momentum tensor operator on the boundary. However, the Weyl connection does not source any current since it comes from a pure gauge mode of the bulk metric. Despite being pure gauge, the appearance of the Weyl connection on the boundary is far from innocuous since it makes the geometric quantities on the boundary Weyl-covariant. Specifically, we will show that the obstruction tensors in the WFG gauge are promoted to Weyl-obstruction tensors, which will play an important role in the construction of the Weyl anomaly in this paper.
\par
The Weyl anomaly is reflected by the nonvanishing trace of the energy-momentum tensor in even dimensions, which has been computed for various conformal field theories\cite{Capper1974,Deser1976,Duff1977,Polyakov1981,Fradkin:1983tg,Bonora:1985cq,Osborn1991,Deser1993,Henningson1998,Boulanger:2007st,Boulanger:2007ab}. The results in $2d$ and $4d$ are well-known:
\begin{align}
2d:\langle T^\mu{}_\mu\rangle= -\frac{c}{24\pi} R\,,\qquad4d:\langle T^\mu{}_\mu\rangle= cW^2-aE^{(4)}\,,
\end{align}
where $W^2$ is the contraction of two Weyl tensors, and $E^{(4)}$ is the Euler density in $4d$. In the context of holography, the Weyl anomaly was first suggested in \cite{Witten:1998qj}, and was then calculated from the bulk in \cite{Liu:1998bu} and \cite{Henningson1998}.
For a holographic theory where we have the vacuum Einstein theory in the bulk, one gets $a=c$ in the 4-dimensional boundary theory as a constraint on the central charges. In the FG gauge, after going through the holographic renormalization procedure by adding counterterms to cancel the divergence extracted by the regulator, one finds that the holographic Weyl anomaly in an even dimension corresponds to the logarithmic term in the bulk volume expansion. In mathematical literature this is also referred to as the Q-curvature \cite{Branson1991,Branson1995,Fefferman2001,Fefferman2003} (see \cite{Chang2008} for a short review), which has been studied by means of obstruction tensors and extended obstruction tensors in \cite{graham2005ambient} and \cite{GRAHAM20091956}. Going into the WFG gauge, it was shown in \cite{Ciambelli:2019bzz} using dimensional regularization that the Weyl anomaly in $2k$-dimension can be extracted directly from the variation of the pole term at the $O(z^{2k-d})$-order of the ``bare" on-shell action under the $d\to 2k^-$ limit. This is the method we will use for computing the Weyl anomaly in this work.
\par
Our goal in this paper is to find the holographic Weyl anomaly in higher dimensions using the advantages of the WFG gauge, and organize the results in a form that manifests its general structure. It has been shown in \cite{Ciambelli:2019bzz} that, up to total derivatives, the Weyl anomaly in $2d$ and $4d$ in the WFG gauge has the same form of that in the FG gauge, but now become Weyl-covariant. We generalize these results to $6d$ and $8d$ by calculating the Weyl anomaly explicitly, and we find that the same statement still holds. Furthermore, we discover that by promoting the obstruction tensors in the FG gauge to the Weyl-obstruction tensors in the WFG gauge, one can use them as natural building blocks for the Weyl anomaly. In this way, we will see clearly how  the WFG gauge Weyl-covariantizes the Weyl anomaly without introducing additional nontrivial cocycles. Our results also reveal some interesting clues about the general form of the holographic Weyl anomaly in any dimension.
\par
This paper will be organized as follows. In Section \ref{Sec2} we briefly introduce the obstruction tensors and extended obstruction tensors in the FG gauge and their properties. In Section \ref{Sec3} we review the WFG gauge as the Weyl-covariant modification of the FG gauge, and how the bulk LC connection induces a Weyl connection on the conformal boundary. More details about the Weyl connection and Weyl geometry are given in Appendix \ref{WG}. In Section \ref{Sec4} we  generalize the results of Section \ref{Sec2} to Weyl-obstruction tensors and extended Weyl-obstruction tensors by solving the Einstein equations in the WFG gauge. The expansions of the Einstein equations can be found in Appendix \ref{AppB0}. Using the Weyl-Schouten tensor and extended Weyl-obstruction tensors as building blocks, in Section \ref{Sec5} we will derive the holographic Weyl anomaly in the WFG gauge in $6d$ and $8d$ after we review the results in $2d$ and $4d$. More details of the calculation are provided in Appendix \ref{AppB}. As a consistency check, we also compute the $8d$ holographic Weyl anomaly in the FG gauge using a completely different approach---the dilatation operator method\cite{Papadimitriou:2004ap,Anastasiou:2020zwc}---which will be presented in Appendix \ref{AppC}. The result agrees with what we get in Section \ref{Sec5}. The expressions for the holographic Weyl anomaly up to $8d$ also suggest the pattern in any dimension, which will be discussed in the end of Section 5. In Section \ref{Sec6} we discuss some aspects of the Weyl structure observed from the role it plays in the formulas for the Weyl-obstruction tensors and Weyl anomaly that we derived. Finally, in Section \ref{Sec7} we summarize our results and point out possible directions for future studies.

\section{Obstruction Tensors}
\label{Sec2}
The obstruction tensor is known as the only irreducible conformal covariant tensor besides the Weyl tensor in an even-dimensional spacetime. The general references for the obstruction tensor are \cite{graham2005ambient,Fefferman2011}, where it was defined precisely in terms of the ambient metric. Instead of providing the formal definition, we will derive the obstruction tensors explicitly in the FG gauge for up to $d=6$ by solving the bulk equations of motion order by order. The same method will also be used in Section \ref{Sec4} for the Weyl-obstruction tensor. 
\par
According to the Fefferman-Graham theorem \cite{Feffe} the metric of a $(d+1)$-dimensional asymptotically locally AdS (AlAdS) spacetime can always be expressed in the following form
\begin{equation}
\label{FG}
\td s^2=L^2\frac{\td z^2}{z^2}+h_{\mu\nu}(z;x)\td x^\mu \td x^\nu\,,\qquad\mu,\nu=0,\cdots,d-1\,,
\end{equation}
where the coordinate $z$ can be considered as a ``radial" coordinate, and $z=0$ is the ``location" of the conformal boundary. When $h_{\mu\nu}=L^2 \eta_{\mu\nu}/z^2 $ with $\eta_{\mu\nu}$ the flat metric, this represents the Poincar\'e metric for $AdS_{d+1}$ spacetime. Near the conformal boundary, $h_{\mu\nu}$ can be expanded with respect to $z$ as follows\cite{Ciambelli:2019bzz}:
\begin{equation}
\label{hex0}
h_{\mu\nu}(z;x)=\frac{L^2}{z^2}\left[\gamma^{(0)}_{\mu\nu}(x)+\frac{z^2}{L^2}\gamma^{(2)}_{\mu\nu}(x)+\cdots\right]+\frac{z^{d-2}}{L^{d-2}}\left[\pi^{(0)}_{\mu\nu}(x)+\frac{z^2}{L^2}\pi^{(2)}_{\mu\nu}(x)+\cdots\right]\,.
\end{equation}
As we mentioned in Section \ref{Intro}, the conformal boundary carries a conformal class of metrics. In the FG expansion $\gamma^{(0)}_{\mu\nu}$ serves as the ``canonical" representative of the conformal class sourcing the energy-momentum tensor of the dual field theory on the boundary, while $\pi^{(0)}_{\mu\nu}$ corresponds to the expectation value of the energy-momentum tensor\cite{Leigh}. Once $\gamma^{(0)}_{\mu\nu}$ is given, each term in the first series can be determined by solving the vacuum Einstein equations with negative cosmological constant in the bulk. Similarly, once $\pi^{(0)}_{\mu\nu}$ is given, the second series will be determined. However, $\pi^{(0)}_{\mu\nu}$ is not completely arbitrary but is actually constrained by the Einstein equations. To be more specific, the $zz$-component of the Einstein equations tells us that $\pi^{(0)}_{\mu\nu}$ is traceless while the $z\mu$-components indicate that it is also divergence-free.
\par
Nevertheless, subtleties will arise when the boundary dimension $d$ is an even integer, since the two series in \eqref{hex0} mix into one. To resolve this issue for an even $d=2k$, we treat $d$ formally as a variable $d\in \mathbb{C}$ in the expansion \eqref{hex0} and let $d$ approach $2k$ from below. As we will see explicitly, when the Einstein equations are satisfied, $\gamma^{(2k)}_{\mu\nu}$ has a first order pole at $d=2k$. For any integer $k\geqslant2$, up to some factor, the coefficient of the pole term (which is actually a meromorphic function of the boundary dimension) is what we define as the \emph{obstruction tensor}, denoted by $\mathcal{O}^{(2k)}_{\mu\nu}$:
\begin{equation}\label{gamma2k}
\gamma^{(2k)}_{\mu\nu}= \frac{c_{(2k)}}{d-2k}\mathcal{O}^{(2k)}_{\mu\nu}+ \tilde{\gamma}_{\mu\nu}^{(2k)}\,,\qquad c_{(2k)}=-\frac{L^{2k}}{2^{2k-3}k!}\frac{\Gamma(d/2-k+1)}{ \Gamma(d/2-1)}\,,
\end{equation}
where the normalization factor $c_{(2k)}$ has been chosen so that the obstruction tensor agrees with the convention of \cite{Fefferman2011}, and the tensor $\tilde{\gamma}^{(2k)}_{\mu\nu}$ is analytic at $d=2k$. 
\par
 Besides holographic dimensional regularization \cite{Leigh}, another common approach is to introduce a logarithmic term for $d=2k$\cite{Henningson1998}, which turns out to be proportional to the obstruction tensor. This is also the origin of the name obstruction tensor, as it obstructs the existence of a formal power series expansion. Note that the tensor $\mathcal{O}^{(2k)}_{\mu\nu}$ is well-defined in any dimension, but only behaves as an ``obstruction" when $d=2k$. The relation between the two approaches will be cleared up at the end of this section once we show how to correctly take the limit for an even $d$ in holographic dimensional regularization. 
\par
Now we present the obstruction tensors explicitly. First, by solving the bulk Einstein equations to the $O(z^2)$-order one finds that
\begin{align}
\label{gamma2}
\frac{\gamma^{(2)}_{\mu\nu}}{L^2}=-\frac{1}{d-2}\left(R^{(0)}_{\mu\nu}-\frac{R^{(0)}}{2(d-1)}\gamma_{\mu\nu}^{(0)}\right)\,,
\end{align}
where $R^{(0)}_{\mu\nu}$ and $R^{(0)}$ represent the Ricci tensor and Ricci scalar of $\gamma_{\mu\nu}^{(0)}$ at the boundary, respectively. One can recognize $\gamma^{(2)}_{\mu\nu}/L^2$ as the Schouten tensor $P_{\mu\nu}$ at the boundary (with a minus sign):
\begin{align}
\label{P}
P_{\mu\nu}&=\frac{1}{d-2}\left(R^{(0)}_{\mu\nu}-\frac{R^{(0)}}{2(d-1)}\gamma_{\mu\nu}^{(0)}\right)\,.
\end{align}
Indeed we notice that there is a first order pole when $d=2$ as expected. However, it is easy to see that  the residue of the pole vanishes identically for $d=2$. This is the reason why it is often stated that there is no obstruction tensor for $d=2$.
\par
At the $O(z^4)$-order, the Einstein equations give us
\begin{align}
\label{gamma4}
\frac{\gamma^{(4)}_{\mu\nu}}{L^4}=-\frac{1}{4(d-4)}B_{\mu\nu}+\frac{1}{4}P_{\rho\mu}P^{\rho}{}_{\nu}\,.
\end{align}
Note that on the boundary, the tensor indices are lowered and raised using $\gamma^{(0)}_{\mu\nu}$ and its inverse $\gamma_{(0)}^{\mu\nu}$. The tensor $B_{\mu\nu}$ is the Bach tensor, which is defined as
\begin{align}
\label{B}
B_{\mu\nu}&=\nabla_{(0)}^\lambda\nabla^{(0)}_\lambda P_{\mu\nu}-\nabla_{(0)}^\lambda\nabla^{(0)}_{\nu} P_{\mu\lambda}-W^{(0)}_{\rho\nu\mu\lambda}P^{\lambda\rho}\,,
\end{align}
where $\nabla^{(0)}_\mu$ is the derivative operator on the boundary associated with $\gamma^{(0)}_{\mu\nu}$, and $W^{(0)}_{\rho\mu\nu\lambda}$ is the Weyl tensor of $\gamma_{\mu\nu}^{(0)}$. We notice that the first term has a pole at $d=4$ and it follows from \eqref{gamma2k} that the obstruction tensor for $d=4$ is just the Bach tensor, i.e.\ $
{\cal O}^{(4)}_{\mu\nu}= B_{\mu\nu}$.\par
Similarly, if we move on to the $O(z^6)$-order of the Einstein equations, we find that $\gamma^{(6)}_{\mu\nu}$ has a pole at $d=6$ and can be written as
\begin{align}
\label{gamma6}
\frac{\gamma^{(6)}_{\mu\nu}}{L^6}&=-\frac{1}{24(d-6)(d-4)}{\cal O}^{(6)}_{\mu\nu}+\frac{1}{6(d-4)}B_{\rho\mu}P^\rho{}_\nu\,.
\end{align}
From \eqref{gamma2k} one can see that ${\cal O}^{(6)}_{\mu\nu}$ is the obstruction tensor for $d=6$, now given by
\begin{align}
\label{O6}
{\cal O}^{(6)}_{\mu\nu}={}&\nabla_{(0)}^\lambda\nabla^{(0)}_\lambda B_{\mu\nu}-2W^{(0)}_{\rho\nu\mu\lambda}B^{\lambda\rho}-4B_{\mu\nu}P+2(d-4)\big(2P^{\rho\lambda}\nabla^{(0)}_\lambda C_{(\mu\nu)\rho}+\nabla^{(0)}_\lambda PC_{(\mu\nu)}{}^\lambda\nn\\
&\qquad\qquad-C^{\rho}{}_{\mu}{}^{\lambda}C_{\lambda\nu\rho}+ \hat\nabla_{(0)}^\lambda P^\rho{}_{(\mu}C_{\nu)\rho\lambda}-W^{(0)}_{\rho\mu\nu\lambda}P^{\lambda}{}_\sigma P^{\sigma\rho}\big)\,,
\end{align}
where $P\equiv P_{\mu\nu}\gamma_{(0)}^{\mu\nu}$, and $C_{\mu\nu\rho}$ is the Cotton tensor on the boundary defined as
\begin{align}
C_{\mu\nu\rho}=\nabla^{(0)}_\rho P_{\mu\nu}-\nabla^{(0)}_\nu P_{\mu\rho}\,.
\end{align}
\par
Let us make a few remarks on some important properties of the obstruction tensors. First, they are symmetric traceless tensors for any boundary dimension $d$. The traceless condition can be derived from the $zz$-component of the Einstein equations at the $O(z^{2k})$-order. Also, the obstruction tensor ${\cal O}^{(2k)}_{\mu\nu}$ is divergence-free when $d=2k$. For instance, divergence of the Bach tensor gives
\begin{align}
\label{divB}
\nabla_{(0)}^\nu B_{\nu\mu}=(d-4)P^{\nu\rho}C_{\rho\nu\mu}\,.
\end{align}
The divergence of the Bach tensor can be read  from the $O(z^4)$-order of the $z\mu$-component of  Einstein equations. In general, at any $O(z^{2k})$-order one finds that the divergence of ${\cal O}^{(2k)}_{\mu\nu}$ is proportional to $d-2k$ and thus vanishes when $d=2k$. The divergence of ${\cal O}^{(2k)}_{\mu\nu}$ can also be obtained by using the following identity
\begin{align}
\label{divP}
\nabla_{(0)}^\nu P_{\nu\mu}=\nabla^{(0)}_\mu P\,.
\end{align}
This is equivalent to the contracted Bianchi identity at the boundary (see Appendix \ref{WG}), which can also be read from the leading order of the $z\mu$-component of  Einstein equations. Finally, a notable feature of ${\cal O}^{(2k)}_{\mu\nu}$ is that it is Weyl-covariant when $d=2k$ with Weyl weight $2k-2$ (for a proof see \cite{graham2005ambient}).
\par
For convenience, we can also absorb the $d$-dependent factors in $\gamma^{(2k)}_{\mu\nu}$ by introducing Graham's extended obstruction tensor $\Omega^{(k-1)}_{\mu\nu}$ ($k\geqslant 2$):
\begin{align}
\label{extO}
\Omega^{(1)}_{\mu\nu}=-\frac{1}{d-4}B_{\mu\nu}\,,\qquad\Omega^{(2)}_{\mu\nu}=\frac{1}{(d-6)(d-4)}\mathcal O^{(6)}_{\mu\nu}\,,\qquad\cdots
\end{align}
The extended obstruction tensor $\Omega^{(k)}_{\mu\nu}$ was precisely defined in \cite{GRAHAM20091956} in the context of the ambient metric. The general relation between the obstruction tensor and extended obstruction tensor is
\begin{align}
\Omega^{(k)}_{\mu\nu}=\frac{(-1)^{k}}{2^k}\frac{\Gamma(d/2-k-1)}{\Gamma(d/2-1)}{\cal O}^{(2k+2)}_{\mu\nu}\qquad (k\geqslant1)\,.
\end{align}
\par
We finish this section by describing how to get the $d\to 2k^{-}$ limit of the two series in \eqref{hex0} properly. By taking the limit carefully we will recover a logarithmic term in the expansion whose coefficient is exactly the obstruction tensor for $d=2k$, which also justifies the name ``obstruction" as we mentioned before. There are two issues one has to deal with while taking the  $d\to 2k^{-}$ limit. First, as we already noted, $\gamma^{(2k)}_{\mu\nu}$ has a pole at $d-2k$, so it diverges in this limit.  Second, the two series mix since both $\gamma^{(2k)}_{\mu\nu}$ and $\pi^{(0)}_{\mu\nu}$ appear at the same order $O(z^{2(k-1)})$ in \eqref{hex0}, for $d=2k$. To keep the $O(z^{2k})$-order finite we pose that  $\pi_{\mu\nu}^{(0)}$ should also have a pole for $d=2k$ proportional to ${\cal O}^{(2k)}_{\mu\nu}$ so that the divergence in $\gamma^{(2k)}_{\mu\nu}$ gets canceled, i.e.\ we claim that $\pi^{(0)}_{\mu\nu}$ has the following form:
\begin{equation}\label{pi0}
\pi^{(0)}_{\mu\nu}= - \frac{c_{(2k)}}{d-2k}{\cal O}^{(2k)}_{\mu\nu} + \tilde{\pi}^{(0)}_{\mu\nu}\,,
\end{equation}
where $\tilde{\pi}^{(0)}_{\mu\nu}$ is finite at $d=2k$. Substituting back \eqref{pi0} and \eqref{gamma2k} to \eqref{hex0} we get
\begin{equation}
h_{\mu\nu}(z;x)=\sum_{n=0}^{k-1}\gamma_{\mu\nu}^{(2n)}\left(\frac{z}{L}\right)^{2n-2} + \big(\tilde\gamma_{\mu\nu}^{(2k)}+ \tilde{\pi}^{(0)}_{\mu\nu}\big)\left(\frac{z}{L}\right)^{2k-2}-c_{(2k)}\left(\frac{z}{L}\right)^{2k-2}\text{ln}\left(\frac{z}{L}\right){\cal O}^{(2k)}_{\mu\nu} +o\big((z/L)^{d}\big)\,.
\end{equation}
This makes contact with the expansion with a logarithmic term (for an even $d$) presented in the literature, e.g.\ \cite{Henningson1998,deHaro:2000vlm,Skenderis2002}.

\section{Weyl-Fefferman-Graham Gauge}
\label{Sec3}
This section is a brief review of the Weyl-Fefferman-Graham (WFG) formalism established in \cite{Ciambelli:2019bzz}. At the end of this section we introduce the ``Weyl quantities" that will appear in  later sections.
\par
The Fefferman-Graham ansatz \eqref{FG} is quite convenient for calculations, especially in the context of holographic renormalization. In this setup, one can induce a Weyl transformation of the boundary metric by a bulk diffeomorphism, namely the PBH transformation\cite{Imbimbo_2000},
\begin{equation}
z\to z'= z/{\cal{B}}(x)\,,\qquad  x^{\mu}\to x'^{\mu}= x^{\mu}+ \xi^{\mu}(z;x)\, , 
\end{equation}
where $\xi^{\mu}(z;x)$ vanish at the boundary $z=0$. The functions $\xi^{\mu}(z;x)$ can be found (infinitesimally) in terms of ${\cal{B}}(x)$ by the constraint that the form of the FG ansatz is preserved under the transformation. However, under the PBH transformation, the subleading terms in the FG expansion \eqref{hex0} do not transform in a Weyl-covariant way. The source of this complication is the compensating diffeomorphisms $\xi^{\mu}(z;x)$ introduced for preserving the FG ansatz. 
\par
This above-mentioned issue motivated the authors of \cite{Ciambelli:2019bzz} to replace the FG ansatz with
\begin{align}
\label{WFG}
\td s^2=L^{2}\left(\frac{\td z}{z}-a_\mu(z;x) \td x^\mu\right)^2
+h_{\mu\nu}(z;x)\td x^\mu \td x^\nu\,,
\end{align}
which was named the Weyl-Fefferman-Graham ansatz. With the additional Weyl structure $a_{\mu}$ added, the form of the WFG ansatz is now preserved under the Weyl diffeomorphism
\begin{align}
\label{weyl}
z\to z'=z/{\cal B}(x)\,,\qquad x^\mu\to x'^\mu=x^\mu\,.
\end{align}
It is not hard to see that the Weyl diffeomorphism \eqref{weyl} induces the following transformation of the fields $a_{\mu}$ and $h_{\mu\nu}$:
\begin{align}
\label{weylha}
a_\mu(z;x)\to a'_{\mu}(z';x)= a_\mu({\cal B}(x)z';x)-\p_\mu\ln{\cal B}(x)\,,\quad h_{\mu\nu}\to h'_{\mu\nu}(z';x)= h_{\mu\nu}({\cal B}(x)z';x)\,.
\end{align}
Thus, we can now induce a Weyl transformation on the boundary and preserve the form of the metric without introducing the irritating $\xi^{\mu}(z;x)$. Note that according to the FG theorem, any AlAdS spacetime can always be expressed in the FG form, and so \eqref{WFG} can be transformed into \eqref{FG} under a suitable diffeomorphism. This indicates that $a_\mu$ is actually pure gauge in the bulk. Another way of going back to the FG gauge is to simply set  $a_\mu$ to zero; in this perspective, the FG gauge is nothing but a special case of the WFG gauge with a fixed gauge.
\par
The main utility of the WFG gauge is that all the terms (except one) in the $z$-expansions of $h_{\mu\nu}(z;x)$ and $a_{\mu}(z;x)$ transform as Weyl tensors under Weyl diffeomorphisms. To see this, let us expand $h_{\mu\nu}$ and $a_\mu$ near $z=0$:
\begin{align}
\label{hex}
h_{\mu\nu}(z;x)&=\frac{L^2}{z^2}\left[\gamma^{(0)}_{\mu\nu}(x)+\frac{z^2}{L^2}\gamma^{(2)}_{\mu\nu}(x)+\cdots\right]+\frac{z^{d-2}}{L^{d-2}}\left[\pi^{(0)}_{\mu\nu}(x)+\frac{z^2}{L^2}\pi^{(2)}_{\mu\nu}(x)+\cdots\right]\,,\\
\label{aex}
a_{\mu}(z;x)&=\left[a^{(0)}_{\mu}(x)+\frac{z^2}{L^2}a^{(2)}_{\mu}(x)+\cdots\right]
+\frac{z^{d-2}}{L^{d-2}}\left[p^{(0)}_{\mu}(x)+\frac{z^2}{L^2}p^{(2)}_{\mu}(x)+\cdots\right]\,.
\end{align}
In the FG gauge where $a_\mu$ is turned off, the FG expansion only includes \eqref{hex}, and the subleading terms $\gamma^{(2k)}_{\mu\nu}$ in the first series are determined solely by the boundary induced metric $\gamma^{(0)}_{\mu\nu}$ and its derivatives. Now with the extra series \eqref{aex}, $\gamma^{(2k)}_{\mu\nu}$ will also depend on $a^{(0)}_\mu$, $a^{(2)}_\mu$, $a^{(4)}_\mu$, etc. 
Moving on, from the transformations \eqref{weylha} under a Weyl diffeomorphism, one finds the transformation of each term in the expansions \eqref{hex} and \eqref{aex} as follows\cite{Ciambelli:2019bzz}:
\begin{align}
\label{GP1}
\gamma^{(2k)}_{\mu\nu}(x)&\to\gamma^{(2k)}_{\mu\nu}(x){\cal B}(x)^{2k-2}\,,\qquad
\pi^{(k)}_{\mu\nu}(x)\to \pi^{(2k)}_{\mu\nu}(x){\cal B}(x)^{d-2+2k}\,,\\
\label{AP}
a^{(2k)}_{\mu}(x)&\to a^{(2k)}_{\mu}(x){\cal B}(x)^{2k}-\delta_{k,0}\p_\mu\ln{\cal B}(x)\,,\qquad
p^{(2k)}_{\mu}(x)\to p^{(2k)}_{\mu}(x){\cal B}(x)^{d-2+2k}\,.
\end{align}
Indeed, we see that almost all the terms in the expansions transform Weyl-covariantly. The only exception is $a^{(0)}_{\mu}$, which transforms inhomogeneously under Weyl transformation, and thus does not have a definite Weyl weight. All the other terms in the expansions \eqref{hex} and \eqref{aex} can be viewed as tensor fields on the boundary and we can easily read off their Weyl weights from the power of ${\cal B}(x)$  appearing in \eqref{GP1} and \eqref{AP}. 
\par
For a metric in the form of \eqref{WFG} defined on the bulk manifold $M$, one can choose a dual form basis and its corresponding vector basis as follows:
\begin{align}
\label{basis}
\bm e^z&=L\frac{\td z}{z}-La_\mu(z;x)\td x^\mu\,,\qquad \bm e^\mu=\td x^\mu\,,\\
\un e_z&=\frac{z}{L}\un\p_z\equiv \un D_z\,,\qquad \un e_\mu=\un\p_\mu+za_\mu(z;x)\un\p_z\equiv \un D_\mu\,.
\end{align}
Then the tangent space at any point $(z,x^{\mu})\in M$ can be spanned by the basis $\{\un D_z,\un D_\mu\}$, and the basis vectors $\{\un D_\mu\}$ form a $d$-dimensional distribution on $M$ which belongs to the kernel of $\bm e^z$. The Lie brackets of these basis vectors are
\begin{align}
\label{DmDn}
[\un D_\mu,\un D_\nu]=Lf_{\mu\nu}\un D_z\,,\qquad[\un D_z,\un D_\mu]=L\varphi_\mu\un D_z\,,
\end{align}
where $\varphi_\mu\equiv D_za_\mu$ and $f_{\mu\nu}\equiv D_\mu a_\nu-D_\nu a_\mu$ ($D_z$ and $D_\mu$ represent taking the derivatives along $\un e_z$ and $\un e_\mu$). According to the Frobenius theorem, the condition for the distribution spanned by $\{\un D_\mu\}$ to be integrable is that $[\un D_\mu,\un D_\nu]=0$, i.e.\ $f_{\mu\nu}=0$. In this case, this distribution defines a hypersurface. For instance, in the FG gauge where  $a_{\mu}$ is turned off, the distribution $\{\un D_\mu\}$ becomes $\{\un \p_\mu\}$, which generates a foliation of constant-$z$ surfaces. However, $\{\un D_\mu\}$ in the WFG gauge is not necessarily an integrable distribution, and thus one needs to keep in mind that the boundary hypersurface $z=0$ is in general not part of a foliation.
\par 
Suppose $\nabla$ is the Levi-Civita (LC) connection on $M$. One can find the connection coefficients of $\nabla$ in the frame $\{\un D_z,\un D_\mu\}$ from its definition \eqref{conncoef}:
\begin{align}
\nabla_{\un D_\mu}\un D_\nu=\Gamma^\lambda{}_{\mu\nu}\un D_\lambda+\Gamma^z{}_{\mu\nu}\un D_z\,.
\end{align}
The coefficients $\Gamma^\lambda{}_{\mu\nu}$ in the above equation define the induced connection coefficients on the distribution over $M$ spanned by $\{\un D_\mu\}$ (see \cite{munozlecanda2018aspects}). Expanding $\Gamma^\lambda{}_{\mu\nu}$ with respect to $z$, at the leading order one finds that
\begin{align}
 \label{IndWeyl}
\Gamma^{\lambda}_{(0)}{}_{\mu\nu}
 &=\frac{1}{2}\gamma_{(0)}^{\lambda\rho}\big(
 \p_\mu \gamma^{(0)}_{\nu\rho}
 +\p_\nu \gamma^{(0)}_{\mu\rho}-\p_\rho \gamma^{(0)}_{\mu\nu}\big)-\big(a^{(0)}_\mu\delta^\lambda{}_\nu+a^{(0)}_\nu\delta^\lambda{}_\mu+a^{(0)}_\rho\gamma_{(0)}^{\lambda\rho}\gamma^{(0)}_{\mu\nu}\big)\,.
\end{align} 
We can see that \eqref{IndWeyl} gives exactly the connection coefficients of a torsion-free connection with Weyl metricity [see \eqref{WeylLC} in Appendix \ref{WG}, where $A_\mu$ and $g_{\mu\nu}$ correspond to $a^{(0)}_\mu$ and $\gamma^{(0)}_{\mu\nu}$]. That is, on the boundary with $z\to0$ we have a connection $\nabla^{(0)}$ satisfying
\begin{align}
\label{nonmetry}
\nabla^{(0)}_\mu\gamma^{(0)}_{\nu\rho}=2a^{(0)}_\mu\gamma^{(0)}_{\nu\rho}\,.
\end{align}
This indicates that although $a_\mu$ is pure gauge in the bulk, its leading order $a_{\mu}^{(0)}$ serves as a Weyl connection at the conformal boundary. Together with the induced metric $\gamma^{(0)}_{\mu\nu}$, they provide a Weyl geometry at the boundary \cite{Folland:1970}. Under a boundary Weyl transformation
\begin{align}
\label{WT}
\gamma_{\mu\nu}^{(0)} \to {\cal B}(x)^{-2} \gamma^{(0)}_{\mu\nu}\,,\qquad a_{\mu}^{(0)}\to a_{\mu}^{(0)}- \partial_{\mu}\text{ln}{\cal}B(x)\,,
\end{align}
for any tensor $T$ (with indices suppressed) with Weyl weight $w_T$ on the boundary, we have
\begin{align}
T\to B^{w_T}T\,,\qquad(\nabla^{(0)}_\mu T+w_Ta^{(0)}_\mu T)\to B^{w_T}(\nabla^{(0)}_\mu T+w_Ta^{(0)}_\mu T)\,.
\end{align} 
One can also absorb the Weyl connection and define $\hat\nabla^{(0)}$ such that
\begin{align}
\hat\nabla^{(0)}_\mu T\equiv\nabla^{(0)}_\mu T+w_Ta^{(0)}_\mu T\,,
\end{align} 
which renders $\hat\nabla^{(0)}_\mu T$ Weyl-covariant. Particularly, Eq.\ \eqref{nonmetry} indicates that $\hat\nabla^{(0)}$ is metricity-free, which makes it convenient for boundary calculations.
\par
Now that we have the Weyl geometry on the boundary, the geometric quantities there are promoted to the ``Weyl quantities". More precisely, for any geometric quantity constructed by the boundary metric $\gamma^{(0)}_{\mu\nu}$ and the LC connection in the FG case, we now have a Weyl-covariant counterpart of it constructed by $\gamma^{(0)}_{\mu\nu}$, $a^{(0)}_\mu$ and $\hat\nabla^{(0)}$ in the WFG case. For instance, we have the Weyl-Riemann tensor $\hat R^{\mu}{}^{(0)}_{\nu\rho\sigma}$, Weyl-Ricci tensor $\hat R^{(0)}_{\mu\nu}$ and Weyl-Ricci scalar $\hat R^{(0)}$. In addition, $f_{\mu\nu}$ induces on the boundary a tensor $f^{(0)}_{\mu\nu}=\p_\mu a^{(0)}_\nu-\p_\nu a^{(0)}_{\mu}$, namely the curvature of the Weyl connection $a^{(0)}$, which is obviously Weyl-invariant. We can also define the Weyl-Schouten tensor $\hat P_{\mu\nu}$ and Weyl-Cotton tensor $\hat C_{\mu\nu\rho}$ on the boundary as follows:
\begin{align}
\label{WP}
\hat P_{\mu\nu}&=\frac{1}{d-2}\bigg(\hat R^{(0)}_{\mu\nu}-\frac{1}{2(d-1)}\hat R^{(0)}\gamma^{(0)}_{\mu\nu}\bigg)\,,\\
\label{WC}
\hat C_{\mu\nu\rho}&=\hat\nabla^{(0)}_{\rho}\hat P_{\mu\nu}-\hat\nabla^{(0)}_{\nu}\hat P_{\mu\rho}\,.
\end{align}
One should notice that the symmetry of the indices of a ``Weyl quantity" is not necessarily the same as the corresponding quantity defined with the LC connection. For instance, the Weyl-Ricci tensor is not symmetric, with its antisymmetric part $\hat R^{(0)}_{[\mu\nu]}=-(d-2)f^{(0)}_{\mu\nu}/2$, and hence the Weyl-Schouten tensor $\hat P_{\mu\nu}$ also contains an antisymmetric part $\hat P_{[\mu\nu]}=-f^{(0)}_{\mu\nu}/2$. In the next section, we will see that the obstruction tensors also have their Weyl-covariant counterparts. More details of the above Weyl quantities are exhibited in Appendix \ref{WG}.

\section{Weyl-Obstruction Tensors}
\label{Sec4}
In the previous section we saw that the WFG gauge in the bulk induces a Weyl geometry on the boundary. Now we would like to determine the higher order terms in the expansion \eqref{hex} and find the obstruction tensors with the Weyl connection turned on. The method is exactly analogous to that in Section \ref{Sec2} for the FG gauge. By solving the bulk Einstein equations order by order in the WFG gauge, we find that $\gamma^{(2k)}_{\mu\nu}$ still has the same form as \eqref{gamma2k}, except that the obstruction tensor ${\cal O}^{(2k)}_{\mu\nu}$ is now promoted to the \emph{Weyl-obstruction tensor} $\hat {{\cal O}}^{(2k)}_{\mu\nu}$. Unlike ${\cal O}^{(2k)}_{\mu\nu}$, which is only Weyl-covariant in $2k$-dimension, the Weyl-obstruction tensors $\hat {{\cal O}}^{(2k)}_{\mu\nu}$ are Weyl-covariant with a weight $2k-2$ in any dimension; that is, under a Weyl transformation \eqref{WT} it transforms in any $d$ as $\hat {{\cal O}}^{(2k)}_{\mu\nu} \to {\cal B}(x)^{2k-2}\hat {{\cal O}}^{(2k)}_{\mu\nu}$.
\par
In principle, $\gamma^{(2k)}_{\mu\nu}$ at any order can be obtained from the Einstein equations by iteration. In this section, we will show solutions of $\gamma^{(2k)}_{\mu\nu}$ obtained from Einstein equations up to $k=3$, and read off the corresponding Weyl-obstruction tensors from them. Some detailed expansions of  Einstein equations can be found in Appendix \ref{AppB0}. 
\par
First, the leading order of the $\mu\nu$-components of the Einstein equations gives
\begin{align}
\label{g2}
\frac{\gamma^{(2)}_{\mu\nu}}{L^2}&=-\frac{1}{d-2}\bigg(\hat R^{(0)}_{(\mu\nu)}-\frac{1}{2(d-1)}\hat R^{(0)}\gamma^{(0)}_{\mu\nu}\bigg)\,.
\end{align}
We notice that this is the symmetric part of the Weyl-Schouten tensor defined in \eqref{WP} with a minus sign, i.e.\
\begin{align}
\label{Pgf}
\frac{\gamma^{(2)}_{\mu\nu}}{L^2}&=-\hat P_{(\mu\nu)}=-\hat P_{\mu\nu}-\frac{1}{2}f^{(0)}_{\mu\nu}\,.
\end{align}
Similar to the FG gauge, one can check that the residue of the pole in \eqref{g2} vanishes identically when $d=2$. Hence, there is no Weyl-obstruction tensor for $d=2$ and so no logarithmic term will appear in the metric expansion in the $d\to 2^{-}$ limit.
\par
Then, solving the $O(z^2)$-order of the $\mu\nu$-components of the Einstein equations yields
\begin{align}
\label{g4}
\frac{\gamma^{(4)}_{\mu\nu}}{L^4}&=-\frac{1}{4(d-4)}\hat{\cal O}^{(4)}_{\mu\nu}+\frac{1}{4}\hat P^{\rho}{}_{\mu}\hat P_{\rho\nu}-\frac{1}{2L^2}\hat\nabla^{(0)}_{(\mu} a_{\nu)}^{(2)}\,,
\end{align}
where $\hat{\cal O}^{(4)}_{\mu\nu}$ is the Weyl-obstruction tensor for $d=4$, namely the Weyl-Bach tensor $\hat B_{\mu\nu}$, given by
\begin{align}
\hat{\cal O}^{(4)}_{\mu\nu}=\hat B_{\mu\nu}=\hat\nabla^{(0)}_\lambda\hat\nabla_{(0)}^\lambda \hat P_{\mu\nu}-\hat\nabla^{(0)}_\lambda\hat\nabla^{(0)}_{\nu} \hat P_{\mu}{}^{\lambda}-\hat W^{(0)}_{\rho\nu\mu\lambda}\hat P^{\lambda\rho}\,.
\end{align}
If we compare \eqref{g4} with the corresponding result \eqref{gamma4} in the FG case, we see that the form of the expression stays almost the same, with all the LC quantities now being promoted to the corresponding Weyl quantities. Besides, in the WFG gauge $\gamma^{(4)}_{\mu\nu}$ also has an additional term involving $a^{(2)}_{\mu}$, which does not contribute to the pole at $d=4$.
\par
Moving on to the $O(z^4)$-order of the Einstein equations we get
\begin{equation}
\label{g6}
\begin{split}
\frac{\gamma^{(6)}_{\mu\nu}}{L^6}=&-\frac{1}{24(d-6)(d-4)}\hat{\cal O}^{(6)}_{\mu\nu}+\frac{1}{6(d-4)}\hat B_{\rho(\mu}\hat P^\rho{}_{\nu)} -\frac{1}{3L^4}\hat\nabla^{(0)}_{(\mu}a^{(4)}_{\nu)}\\
&-\frac{1}{L^4}a^{(2)}_{\mu}a^{(2)}_{\nu}+\frac{1}{6L^2}a^{(2)}\cdot a^{(2)}\gamma^{(0)}_{\mu\nu}+\frac{1}{6L^2}\hat\nabla^{(0)}_{(\mu}(\hat P^{\rho}{}_{\nu)}a^{(2)}_\rho)+\frac{1}{2L^4}\hat\gamma_{(2)}^{\sigma}{}_{\mu\nu}a^{(2)}_\sigma\, ,
\end{split}
\end{equation}
where $\hat\gamma_{(2)}^{\sigma}{}_{\mu\nu}\equiv-\frac{L^2}{2}(\hat\nabla^{(0)}_\mu\hat P^\sigma{}_{\nu}+\hat\nabla^{(0)}_\nu\hat P_\mu{}^\sigma-\hat\nabla_{(0)}^\sigma\hat P_{\mu\nu})$, and $\hat{\cal O}^{(6)}_{\mu\nu}$ is the Weyl-obstruction tensor for $d=6$:
\begin{equation}
\label{WO6}
\begin{split}
\hat{\cal O}^{(6)}_{\mu\nu}={}&\hat\nabla_{(0)}^\lambda\hat\nabla^{(0)}_\lambda\hat  B_{\mu\nu}-2\hat W^{(0)}_{\rho\nu\mu\lambda}\hat B^{\lambda\rho}-4\hat P\hat B_{\mu\nu}+2\hat P_{\rho(\nu}\hat B^{\rho}{}_{\mu)}-2\hat B^{\rho}{}_{(\mu}\hat P_{\nu)\rho}\\
&+2(d-4)\bigg(\hat\nabla_{(0)}^\lambda\hat  C_{\lambda\rho(\mu}\hat P^{\rho}{}_{\nu)} -\hat P^{\lambda\rho}\hat\nabla^{(0)}_{(\mu}\hat  C_{\nu)\rho\lambda}+2\hat P^{(\rho\lambda)}\hat\nabla^{(0)}_\lambda \hat C_{(\mu\nu)\rho}+\hat\nabla^{(0)}_\lambda\hat P^{\rho\lambda}\hat C_{(\mu\nu)\rho}\\
&\qquad\qquad\qquad-\hat C^{\rho}{}_{\mu}{}^{\lambda}\hat C_{\lambda\nu\rho}+ \hat\nabla_{(0)}^\lambda\hat P^\rho{}_{(\mu}\hat C_{\nu)\rho\lambda}-\hat W^{(0)}_{\rho(\nu\mu)\lambda}\hat P^{\lambda}{}_\sigma\hat P^{\sigma\rho}\bigg)\,.
\end{split}
\end{equation}
It is easy to verify that  \eqref{g6} and \eqref{WO6} go back to the FG expressions \eqref{gamma6} and \eqref{O6} when we turn off the Weyl structure $a_\mu$. Note that when the Weyl connection is turned off, the first term inside the parentheses of \eqref{WO6} vanishes due to \eqref{divC}, and the second term there vanishes since the LC Schouten tensor $\mathring P_{\mu\nu}$ is symmetric. Once again, we observe that all the $a^{(2)}_\mu$ and $a^{(4)}_\mu$ terms that appear in $\gamma^{(6)}_{\mu\nu}$ do not contribute to the pole at $d=6$ and thus are not part of the obstruction tensor $\hat{\cal O}_{\mu\nu}^{(6)}$. We will discuss this more in Section \ref{Sec6}.
\par
Just as ${\cal O}_{\mu\nu}^{(2k)}$ derived in the FG gauge, all the $\hat{\cal O}_{\mu\nu}^{(2k)}$ are also symmetric traceless tensors, and they are divergence-free when $d=2k$. These properties can either be verified by using the result from the $\mu\nu$-components of the Einstein equations (``evolution equations"), or read off from the $zz$- and $z\mu$-components of the Einstein equations (``constraint equations"). More specifically, plugging $\gamma^{(2k)}_{\mu\nu}$ into the $zz$-component of the Einstein equations we can see that $\mathcal{\hat O}^{(2k)}_{\mu\nu}$ is traceless in any dimension, and the same result can also be obtained by taking the trace of the $\mu\nu$-components of the Einstein equations. To see that $\hat{\cal O}_{\mu\nu}^{(2k)}$ is divergence-free when $d=2k$, we can plug $\gamma^{(2k)}_{\mu\nu}$ into the $z\mu$-components of the Einstein equations. For instance, the $O(z^4)$-order of the $z\mu$-equations gives
\begin{align}
\label{divWB}
\hat\nabla_{(0)}^\nu\hat B_{\nu\mu}=(d-4)\hat P^{\nu\rho}(\hat C_{\rho\nu\mu}+\hat C_{\mu\nu\rho})\,,
\end{align}
and so the divergence of $\hat B_{\mu\nu}$ vanishes when $d=4$. In the FG gauge where the Schouten tensor is symmetric, the second term in the bracket vanishes and so \eqref{divWB} goes back to \eqref{divB}. On the other hand, the divergence of $\hat{\cal O}_{\mu\nu}^{(2k)}$ can also be derived from a direct calculation by using repeatedly the Weyl-Bianchi identity
\begin{align}
\label{divWP}
\hat\nabla_{(0)}^\nu\hat P_{\nu\mu}=\hat\nabla^{(0)}_\mu\hat P\,,
\end{align}
which can be read off from the $O(z^2)$-order of the $z\mu$-equation. The above discussion indicates that the $zz$- and $z\mu$-components of the Einstein equations do not contain more information about $\gamma^{(2k)}_{\mu\nu}$ than the $\mu\nu$-components of  Einstein equations. Note that here we only talk about the equations of motion for $\gamma^{(2k)}_{\mu\nu}$. At $O(z^d)$-order the $zz$- and $z\mu$-equations do provide new constraints on $\pi^{(0)}_{\mu\nu}$, while the $\mu\nu$-equations on $\pi^{(0)}_{\mu\nu}$ become trivial.
\par
It is also convenient to define the \emph{extended Weyl-obstruction tensor} $\hat{\Omega}^{(k)}_{\mu\nu}$ as the Weyl-covariant version of the extended obstruction tensor defined in \eqref{extO}. For example, for $k=1$ and $k=2$ we have
\begin{align}
\label{extWO}
\hat\Omega^{(1)}_{\mu\nu}=-\frac{1}{d-4}\hat B_{\mu\nu}\,,\qquad\hat\Omega^{(2)}_{\mu\nu}=\frac{1}{(d-6)(d-4)}\hat{\mathcal O}^{(6)}_{\mu\nu}\,.
\end{align}
\par
Similar to the FG case, the Weyl-obstruction tensor $\hat{\cal O}_{\mu\nu}^{(2k+2)}$ is also proportional to the residue of the extended Weyl-obstruction tensor $\hat{\Omega}^{(k)}_{\mu\nu}$. Both the Weyl-obstruction tensors and the extended Weyl-obstruction tensors can be defined following \cite{graham2005ambient,GRAHAM20091956} by promoting the ambient metric to the ``Weyl-ambient metric". We will discuss this in detail in a separate publication.

\section{Holographic Weyl Anomaly}
\label{Sec5}
\subsection{Weyl-Ward Identity}\label{WeylWardIdentity}
In this section, we first discuss the anomalous Weyl-Ward identity for a general field theory on a background Weyl geometry following \cite{Ciambelli:2019bzz}, and then we focus on holographic theories in the WFG gauge. Later, we will compute the Weyl anomaly for a holographic theory in the WFG gauge up to $d=8$.
\par
Essentially, for a $d$-dimensional field theory\footnote{From now on, we will work in the Euclidean signature. We also adopt natural units where $c=\hbar=1$.} coupled to a background metric $\gamma^{(0)}_{\mu\nu}$ and a Weyl connection $a^{(0)}_{\mu}$, the Weyl anomaly comes from an additional exponential factor arising in the path integral after applying a Weyl transformation:
\begin{align}
\label{Z}
Z[\gamma^{(0)},a^{(0)}]= \E^{-{\cal A}[{\cal B}(x);\gamma^{(0)},a^{(0)}]} Z[\gamma^{(0)}/{\cal B}(x)^{2},a^{(0)}-\text d\ln {\cal B}(x)]\,.
\end{align}
The anomaly ${\cal A}[{\cal B}(x);g,a]$ should satisfy the 1-cocycle condition \cite{Manvelyan:2001pv,BONORA1983305}
\begin{align}
{\cal A} [{\cal B''} {\cal B'};\gamma^{(0)},a^{(0)}]= {\cal A} [{\cal B'};\gamma^{(0)},a^{(0)}] + {\cal A} [{\cal B''}; \gamma^{(0)}/({\cal B'})^{2},a^{(0)}- \td\ln {\cal B'}]\,.
\end{align}
For any non-exact Weyl-invariant $d$-form $\bm A[\gamma_{(0)},a_{(0)}]$, one can check that ${\cal A}[{\cal B}(x);\gamma^{(0)},a^{(0)}]= \int (\ln {\cal B})\bm{A}$ satisfies the cocycle condition, and thus it is a possible candidate for the Weyl anomaly. However, if $\bm A$ is exact, ${\cal A}$ would be cohomologically trivial since it can be written as the difference of a Weyl-transformed local functional. The linearly independent choices of $\bm A$ in non-trivial cocycles correspond to different central charges.
\par
In general, the background fields $\gamma^{(0)}_{\mu\nu}$ and $a^{(0)}_\mu$ are the sources of the energy-momentum tensor operator $T^{\mu\nu}$ and the Weyl current operator $J^\mu$, respectively:
\begin{align}
\langle T^{\mu\nu}(x)\rangle=\frac{2}{\sqrt{-\det\gamma^{(0)}}}\frac{\delta S}{\delta \gamma^{(0)}_{\mu\nu}(x)}\,,\qquad \langle J^\mu(x)\rangle=-\frac{1}{\sqrt{-\det\gamma^{(0)}}}\frac{\delta S}{\delta a^{(0)}_\mu(x)}\,.
\end{align}
Expanding the quantum effective action $S\equiv-\ln Z$ to the first order under an infinitesimal Weyl transformation and integrating by parts, for a theory with a Weyl anomaly we obtain
\begin{align}
\label{QFTWeylWard}
\frac{1}{\sqrt{-\det\gamma^{(0)}}}\frac{\delta \cal A}{\delta \ln{\cal B}(x)}=\big\langle T^{\mu\nu}(x)\gamma^{(0)}_{\mu\nu}(x)+\hat{\nabla}^{(0)}_\mu  J^\mu(x)\big\rangle\,.
\end{align}
This is the (anomalous) Weyl-Ward identity. As we can see, besides the trace of the energy-momentum tensor that appears in the usual case, the divergence of the Weyl current also contributes to the Ward identity when the Weyl connection is turned on.
\par
Let us now focus on a holographic field theory dual to the vacuum Einstein theory in the $(d+1)$-dimensional bulk. The holographic dictionary provides the relation between the on-shell classical bulk action $S_{bulk}$ and quantum effective action $S_{bdr}$ of the field theory on the boundary \cite{Witten:1998qj}:
\begin{align}
\label{dict}
\exp\left(-S_{bulk}[g;\gamma_{(0)},a_{(0)}]\right)=\exp\left(-S_{bdr}[\gamma_{(0)},a_{(0)}]\right)\,,
\end{align}
where $\gamma_{(0)}$ and $a_{(0)}$ are the boundary values of $h$ and $a$ as shown in \eqref{hex} and \eqref{aex}. Since $a_{\mu}$ is pure gauge in the bulk, $a^{(0)}_\mu$ could be gauged away and hence it is not expected to source any current on the boundary. The role of the $a^{(0)}_{\mu}$, however, is important since it makes the energy-momentum tensor along with all the geometric quantities on the boundary Weyl-covariant. On the other hand, the $p^{(0)}_\mu$ also plays a role in the Weyl-Ward identity. In the FG gauge, $\pi^{(0)}_{\mu\nu}$ corresponds to the expectation value of $T_{\mu\nu}$; the Ward identity for the Weyl symmetry shows that the trace of $\pi^{(0)}_{\mu\nu}$ vanishes, which can be read off from the $O(z^d)$-order of the $zz$-component of the Einstein equations\cite{Leigh}. In the WFG gauge, this equation now gives
\begin{align}
\label{boundaryWI}
0=\frac{d}{2L^2}\gamma_{(0)}^{\mu\nu}\pi^{(0)}_{\mu\nu}+\hat\nabla^{(0)}\cdot p_{(0)}\,.
\end{align}
Besides $\pi^{(0)}_{\mu\nu}$, there is an additional term $\hat\nabla^{(0)}\cdot p_{(0)}$ which represents a gauge ambiguity of $a_\mu$. This suggests that the energy-momentum tensor in the WFG gauge acquires an extra piece, which now can be considered as an ``improved" energy-momentum tensor $\tilde T_{\mu\nu}$ (\`a la\cite{BELINFANTE1940449,Callan1970}):
\begin{equation}
\label{improvedT}
\langle \kappa^2\tilde   T_{\mu\nu}\rangle=\frac{d}{2 L^2}\pi^{(0)}_{\mu\nu}+ \hat\nabla^{(0)}_{(\mu} p^{(0)}_{\nu)}\,,
\end{equation}
where $\kappa^2 =8\pi G$.\footnote{The energy-momentum tensor \eqref{improvedT} in the WFG gauge can be verified using the prescription introduced in\cite{deHaro:2000vlm}.}
It is easy to see that the trace of this energy-momentum tensor gives the right-hand side of \eqref{boundaryWI}. One can also find that the $z\mu$-components of the Einstein equations at the $O(z^d)$-order give exactly the conservation law $\langle \hat\nabla_{(0)}^\mu\tilde T_{\mu\nu}\rangle =0$ [see \eqref{emz}], which is the Ward identity corresponding to the boundary diffeomorphisms. Therefore, in the holographic case we can write the anomalous Weyl-Ward identity \eqref{QFTWeylWard} as
\begin{equation}
\label{holoWeylWard}
\frac{1}{\sqrt{-\det\gamma^{(0)}}}\frac{\delta \cal A}{\delta \ln{\cal B}(x)}=\big\langle\tilde T^{\mu\nu}(x)\gamma^{(0)}_{\mu\nu}(x)\big\rangle\,.
\end{equation}
Notice that one should distinguish $p^{(0)}_\mu$ and the Weyl current $J_\mu$. Unlike $\pi_{\mu\nu}^{(0)}$ which is sourced by $\gamma^{(0)}_{\mu\nu}$, $p^{(0)}_\mu$ is not sourced by $a^{(0)}_\mu$ since $a_\mu$ is pure gauge in the bulk. In the boundary field theory, the Weyl current  $J_{\mu}$ vanishes identically, while $p^{(0)}_{\mu}$ contributes to the expectation value of $\tilde T_{\mu\nu}$ as an ``improvement". In a generic non-holographic field theory defined on the background with Weyl geometry, there may exist a nonvanishing $J_{\mu}$ sourced by the Weyl connection $a^{(0)}_\mu$ (see \cite{Ciambelli:2019bzz} for an example).
\par
Using the basis $\{\bm e^z,\bm e^\mu=\td x^\mu\}$ in \eqref{basis}, the bulk on-shell Einstein-Hilbert action with negative cosmological constant can be written as
\begin{align}
\label{SEH}
S_{bulk}&=\frac{1}{2\kappa^2}\int_M\sqrt{-\det g}\,(R-2\Lambda)\bm{e}^z\wedge \td x^1\wedge\cdots\wedge \td x^d\,.
\end{align}
Note that $\sqrt{-\det g}=\sqrt{-\det h}$. Considering the vacuum Einstein equation in the bulk and the expansion
\begin{align}
\label{sqrth}
\sqrt{-\det h}&=\left(\frac{L}{z}\right)^d\sqrt{-\det\gamma^{(0)}}\left(1+\frac{1}{2}\left(\frac{z}{L}\right)^2 X^{(1)}+\frac{1}{2}\left(\frac{z}{L}\right)^4X^{(2)}+\cdots+\frac{1}{2}\left(\frac{z}{L}\right)^dY^{(1)}+\cdots\right)\,,
\end{align}
one can expand \eqref{SEH} as
\begin{align}
\label{Sbulk}
S_{bulk}&=-\frac{L^{-2}}{\kappa^2}\int_M \left(\frac{L}{z}\right)^d\left(d+\frac{d}{2}\left(\frac{z}{L}\right)^2 X^{(1)}+\frac{d}{2}\left(\frac{z}{L}\right)^4X^{(2)}+\cdots+\frac{d}{2}\left(\frac{z}{L}\right)^dY^{(1)}+\cdots\right)\bm{e}^z\wedge vol_\Sigma\,,
\end{align}
where $vol_\Sigma\equiv \sqrt{-\det\gamma^{(0)}}\td x^1\wedge\cdots\wedge\td x^d$. 
\par
When the bulk action transforms under a Weyl diffeomorphism, the corresponding boundary theory undergoes a Weyl transformation. However, the diffeomorphism invariance of the bulk Einstein theory does not imply the Weyl invariance on the boundary when there is an anomaly\cite{Mazur2001}, since it follows from \eqref{Z} that
\begin{align}
\label{Weyltrans}
0=S_{bulk}[g|z',x']-S_{bulk}[g|z,x]=S_{bdr}[\gamma'_{(0)},a'_{(0)}|x]-S_{bdr}[\gamma_{(0)},a_{(0)}|x]+ {\cal A}[{\cal B}]\,,
\end{align}
where $(z',x')=(z/{\cal B},x)$ for the bulk and $\gamma'_{(0)}=\gamma_{(0)}/{\cal B}^2$, $a'_{(0)}=a_{(0)}-\td\ln{\cal B}$ for the boundary.
\par
Normally, to compute the Weyl anomaly first one needs to regularize the bulk on-shell action \eqref{Sbulk} by introducing a cutoff surface at some small value of $z=\epsilon$, and then add counterterms to cancel the divergences when $\epsilon\to 0$\cite{Henningson1998}. This is essentially how the Weyl anomaly arises since the regulator breaks the Weyl symmetry and causes the appearance of a logarithmically divergent term. However, since we do not assume that we have an integrable distribution when the Weyl structure is turned on, the cutoff regularization scheme is inconvenient for the WFG gauge. It has been elucidated in \cite{Ciambelli:2019bzz} using dimensional regularization that the Weyl anomaly can be extracted from the pole of $S_{bulk}$ that arises in an even dimension. By evaluating the difference of the pole term in $S_{bulk}$ under a Weyl diffeomorphism, one finds that the Weyl anomaly ${\cal A}_k$ of the $2k$-dimensional boundary theory is
\begin{align}
\label{Ak}
{\cal A}_k=\frac{k}{\kappa^2L}\int\ln {\cal B} X^{(k)}_{d=2k}vol_\Sigma\,.
\end{align}
Therefore, to find the Weyl anomaly in $2k$-dimension, we only have to compute $X^{(k)}$ coming from the expansion of $\sqrt{-\det h}$.

\subsection{Weyl Anomaly in $2d$ and $4d$}
Now let us apply \eqref{Ak} to $2d$ and $4d$. Here we first go over the WFG results presented in \cite{Ciambelli:2019bzz}, and then make a few important remarks. To find the holographic Weyl anomaly in $2d$ and $4d$ all we have to do is plug in the expressions of $X^{(1)}$ and $X^{(2)}$ obtained from the $zz$-components of the Einstein equations (see Appendix \ref{AppB0}), that is,
\begin{align}
\label{2d4d}
X^{(1)} =-\frac{L^2}{2(d-1)}\hat R\,,\qquad X^{(2)}=-\frac{L^4}{4(d-2)^2}\bigg(\hat R_{\mu\nu}\hat R^{\nu\mu}-\frac{d}{4(d-1)}\hat R^2\bigg)-\frac{L^2}{2}\hat\nabla\cdot a^{(2)}\,.
\end{align}
[From now on we will drop the label ``(0)" for the boundary curvature quantities and derivative operator when there is no confusion.] First we look at the Weyl anomaly in $d=2$:
\begin{align}
\label{2dA}
{\cal A}_1&=\frac{1}{\kappa^2L}\int\ln {\cal B} X^{(1)}_{d=2}vol_\Sigma =-\frac{L}{16\pi G}\int \ln {\cal B}\hat R\sqrt{-\det\gamma^{(0)}}\td^2x\,,
\end{align}
where in the second equality we used \eqref{2d4d}. Then, it follows from \eqref{holoWeylWard} that the Weyl-Ward identity now reads
\begin{align}
\langle{\tilde T}^\mu{}_\mu\rangle=-\frac{L}{16\pi G}\hat R\,.
\end{align}
We can see that the right-hand side of this result has exactly the same form as what we get from the standard calculation in the FG gauge, except that the curvature scalar now is Weyl-covariant. Similarly, plugging \eqref{2d4d} into \eqref{Ak}, we find that the Weyl anomaly in $d=4$ can be written as
\begin{align}
\label{4dA}
{\cal A}_2&=\frac{2}{\kappa^2L}\int\ln {\cal B} X^{(2)}_{d=4}vol_\Sigma=-\frac{L}{8\pi G}\int\bigg[\frac{L^2}{8}\Big(\hat R_{\mu\nu}\hat R^{\nu\mu}-\frac{1}{3}\hat R^2\Big)+\hat\nabla\cdot a^{(2)}\bigg]\ln {\cal B}\sqrt{-\det\gamma^{(0)}}\td^4x\,.
\end{align}
Again, one can immediately tell that the right-hand side of this result matches the standard FG result (e.g. \cite{Henningson1998}) if we turn off the Weyl structure. 
\par
There are a few things worth paying attention to: first, in the $2d$ Weyl anomaly \eqref{2dA}, the Weyl-Ricci scalar is also the Weyl-Euler density $E^{(2)}$ in $2d$, i.e.\ the Euler density Weyl-covariantized by the Weyl connection. Furthermore, we can rewrite the $4d$ Weyl anomaly \eqref{4dA} as
\begin{align}
\label{A2}
{\cal A}_2&=-\frac{L}{8\pi G}\int\bigg[\frac{L^2}{16}\Big(\hat W_{\mu\nu\rho\sigma}\hat W^{\rho\sigma\mu\nu}-\hat E^{(4)}\Big)+\hat\nabla\cdot a^{(2)}\bigg]\ln {\cal B}\sqrt{-\det\gamma^{(0)}}\td^4x\,,
\end{align}
where $\hat E^{(4)}$ is the Weyl-Euler density in $4d$:
\begin{align}
\hat E^{(4)}=\hat R_{\mu\nu\rho\sigma}\hat R^{\rho\sigma\mu\nu}-4\hat R_{\mu\nu}\hat R^{\nu\mu}+\hat R^2\,.
\end{align}
Traditionally, the Euler density $E^{(2k)}$ without the Weyl connection is called the type A Weyl anomaly, which is topological in $2k$-dimension and not Weyl-invariant, while the type B Weyl anomaly is the Weyl-invariant part of the anomaly \cite{Deser1993}. Here we find that in the WFG gauge, this classification of the Weyl anomaly is still available, with the Weyl-Euler density now Weyl-invariant since the curvature quantities in this setup are endowed with Weyl covariance.
\par
Also, notice that the subleading term $a^{(2)}_\mu$ of $a_\mu$ only makes an appearance in the anomaly through a cohomologically trivial term, i.e.\ we can express it as a Weyl-transformed local functional as follows:
\begin{align}
\int\td^4x\sqrt{-\det\gamma_{(0)}}\ln {\cal B}\,\hat\nabla_\mu a_{(2)}^\mu=\int\td^4x\sqrt{-\det\gamma'_{(0)}}\,a'^{(0)}_\mu a'^\mu_{(2)}-\int\td^4x\sqrt{-\det\gamma_{(0)}}\,a^{(0)}_\mu a_{(2)}^\mu\,,
\end{align}
where $a'^\mu_{(2)}={\cal B}^4a_{(2)}^\mu$, and the boundary term due to integrating by parts is ignored. We will see that this is a generic feature of the Weyl anomaly in the WFG gauge for any dimension.
\par
Although in \eqref{2dA} and \eqref{4dA} we expressed the holographic Weyl anomaly in $2d$ and $4d$ in terms of curvature to match the corresponding familiar results in the FG gauge, we can also express them alternatively in terms of the Weyl-Schouten tensor:
\begin{align}
\label{X1X2}
\frac{X^{(1)}}{L^2}=-\hat P\,,\qquad \frac{X^{(2)}}{L^4}=-\frac{1}{4}\tr(\hat P^2)+\frac{1}{4}\hat P^2-\frac{1}{2L^2}\hat\nabla\cdot a^{(2)}\,.
\end{align}
Then \eqref{2dA} and \eqref{4dA} can be written as
\begin{align}
\label{2dAP}
{\cal A}_1&=-\frac{L}{\kappa^2}\int\td^2x\sqrt{-\det\gamma^{(0)}}\ln {\cal B} \hat P\,,\\
\label{4dAP}
{\cal A}_2&=-\frac{L^3}{\kappa^2}\int\td^4x\sqrt{-\det\gamma^{(0)}}\ln {\cal B} \bigg(\frac{1}{2}\tr(\hat P^2)-\frac{1}{2}\hat P^2+\frac{1}{L^2}\hat\nabla\cdot a^{(2)}\bigg)\,.
\end{align}
In higher dimensions, $X^{(k)}$ can be expressed in terms of $\gamma_{\mu\nu}^{(0\leqslant j\leqslant2k)}$ (see Appendix \ref{AppB}). By solving the Einstein equations we have seen that these terms can all be expressed in terms of $\hat P_{\mu\nu}$ and $\hat{\cal O}^{(2<j<2k)}_{\mu\nu}$. Therefore, we will use the Weyl-Schouten tensor and Weyl-obstruction tensors as the building blocks for the Weyl anomaly in even dimensions.

\subsection{Weyl Anomaly in $6d$}\label{Weyl6d}
After revisiting the results in $2d$ and $4d$, we will now present our computations for $6d$ and $8d$. In principle, $X^{(k)}$ can be obtained by solving Einstein equations as we have done for $2d$ and $4d$. However, as the dimension goes higher, computing the curvature will become extremely tedious. To facilitate the computation in higher dimensions, we can use a more efficient way of organizing the Einstein equations which helps us avoid the curvature tensors, namely to use the Raychaudhuri equation of the congruence generated by $\un D_z$. The details of the Raychaudhuri equation and its expansions are given in Appendix \ref{AppB}.
\par
To solve for $X^{(3)}$, we need to expand $\sqrt{-\det h}$ to the order $O(z^{6-d})$.  Using \eqref{Ray6d} and plugging the results we have got for $\gamma^{(2)}_{\mu\nu},\gamma^{(4)}_{\mu\nu}$ and $X^{(1)},X^{(2)}$ into \eqref{X3}, we obtain
\begin{align}
\frac{X^{(3)}}{L^6}=&-\frac{1}{12}\tr(\hat P^3)+\frac{1}{8}\tr(\hat P^2)\hat P-\frac{1}{24}\hat P^3+\frac{1}{12}\tr(\hat\Omega^{(1)}\hat P)\nn\\
\label{X3P}
&+\frac{1}{6L^4}(d-6)a^2_{(2)}-\frac{1}{3L^4}\hat\nabla\cdot a^{(4)}-\frac{1}{12L^2}\hat\nabla_\mu\big[a^{(2)}_\nu(3\hat P^{\mu\nu}+\hat P^{\nu\mu}
-3\hat P \gamma_{(0)}^{\mu\nu})\big]\,,
\end{align}
where we used the extended Weyl-obstruction tensor $\hat\Omega^{(1)}_{\mu\nu}$ defined in \eqref{extWO}. Notice first that the $a^{(2)}_\mu$ quadratic term in $X^{(3)}$ vanishes in $6d$, and thus does not contribute to the Weyl anomaly. Then, it follows from \eqref{Ak} that the Weyl anomaly in $6d$ is
\begin{align}
\label{6dA}
{\cal A}_3={}&\frac{3}{\kappa^2L}\int\ln {\cal B} X^{(3)}_{d=6}vol_\Sigma\nn\\
={}&-\frac{L^5}{\kappa^2}\int \td^6x\sqrt{-\det\gamma^{(0)}}\ln {\cal B} \bigg(\frac{1}{4}\tr(\hat P^3)-\frac{3}{8}\tr(\hat P^2)\hat P+\frac{1}{8}\hat P^3-\frac{1}{4}\tr(\hat\Omega^{(1)}\hat P)\nn\\
&+\frac{1}{L^4}\hat\nabla\cdot a^{(4)}+\frac{1}{4L^2}\hat\nabla_\mu\big[a^{(2)}_\nu(3\hat P^{\mu\nu}+\hat P^{\nu\mu}
-3\hat P \gamma_{(0)}^{\mu\nu})\big]\bigg)\,.
\end{align}
Just as what we have shown for the $4d$ case, the subleading terms in the expansion of $a_\mu$ appear only in total derivatives and thus only contribute to cohomologically trivial terms in the $6d$ Weyl anomaly. When we turn off $a_\mu^{(0)}$ and $a_\mu^{(2)}$, this result agrees with the holographic Weyl anomaly in the FG gauge computed in \cite{Henningson1998}. \par
Usually, the Weyl anomaly in $6d$ is written as a linear combination of the $6d$ Euler density and three conformal invariants in $6d$ (see \cite{Bonora:1985cq,Deser1993,Henningson1998}), which represents the four central charges in $6d$. The result we obtained can also be written in this way, which means the classification of type A and type B anomalies still holds for the WFG gauge in $6d$. However, as we will discuss shortly, the expression we have in \eqref{X3P} in terms of $\hat P_{\mu\nu}$ and $\hat\Omega^{(1)}_{\mu\nu}$ reveals some interesting aspects of the Weyl anomaly.

\subsection{Weyl Anomaly in $8d$}\label{Weyl8d}
Expanding $\sqrt{-\det h}$ to the order $O(z^{8-d})$, we have $X^{(4)}$ in \eqref{X4}. Using \eqref{Ray8d} and plugging the results up to $\gamma^{(6)}_{\mu\nu}$ and $X^{(3)}$ into \eqref{X4}, we have
\begin{align}
\label{X4P}
\frac{X^{(4)}}{L^8}=&-\frac{1}{32}\tr(\hat P^4)+\frac{1}{24}\tr(\hat P^3)\hat P+\frac{1}{64}(\tr (\hat P^2))^2-\frac{1}{32}\tr(\hat P^2)\hat P^2+\frac{1}{192}\hat P^4\nn\\
&-\frac{1}{24}\tr(\hat\Omega^{(1)}\hat P)\hat P+\frac{1}{24}\tr(\hat\Omega^{(1)}\hat P^2)-\frac{1}{96}\tr(\hat\Omega^{(1)}\hat\Omega^{(1)})-\frac{1}{96}\tr(\hat\Omega^{(2)}\hat P)\nn\\
&+\frac{d-8}{4L^6}a^{(4)}\cdot a^{(2)}+\frac{d-8}{12L^4}a^{(2)}_\mu a^{(2)}_\nu(\hat P^{\mu\nu}-\hat P\gamma_{(0)}^{\mu\nu})+\text{total derivatives}\,.
\end{align}
As expected, all the terms in \eqref{X4P} that involve $a^{(2)}_\mu$, $a^{(4)}_\mu$, $a^{(6)}_\mu$ either vanish when $d=8$ or contribute only to the total derivatives. The details of the total derivatives are given in \eqref{X8t}. Plugging \eqref{X4P} into \eqref{Ak}, we obtain the holographic Weyl anomaly in $8d$:

\begin{align}
\label{8dA}
{\cal A}_4={}&\frac{4}{\kappa^2L}\int\ln {\cal B} X^{(4)}_{d=8}vol_\Sigma\nn\\
={}&-\frac{L^7}{\kappa^2}\int\td^8x\sqrt{-\det\gamma^{(0)}}\ln {\cal B} \bigg(\frac{1}{8}\tr(\hat P^4)-\frac{1}{6}\tr(\hat P^3)\hat P-\frac{1}{16}(\tr (\hat P^2))^2+\frac{1}{8}\tr(\hat P^2)\hat P^2-\frac{1}{48}\hat P^4\nn\\
&+\frac{1}{6}\tr(\hat\Omega^{(1)}\hat P)\hat P-\frac{1}{6}\tr(\hat\Omega^{(1)}\hat P^2)+\frac{1}{24}\tr(\hat\Omega^{(1)}\hat\Omega^{(1)})+\frac{1}{24}\tr(\hat\Omega^{(2)}\hat P)+\text{total derivatives}\bigg)\,.
\end{align}
Once again, we can see that the subleading terms in $a_\mu$ only have cohomologically trivial contributions. If we go back to the FG gauge, then this result agrees with the renormalized volume coefficient for $k=4$ shown in \cite{GRAHAM20091956}. One can also write the FG version of the above result in the traditional way as a linear combination of the type A and type B anomalies, i.e.\ the Euler density and Weyl invariants (the list of Weyl invariants in $8d$ can be found in \cite{Boulanger:2004zf}). We naturally expect that this classification can also be applied to the holographic Weyl anomaly in the WFG gauge for higher dimensions.

\subsection{Building Blocks of the Weyl Anomaly}
As we have seen, if we ignore the total derivatives that depend on the subleading terms of the $a_\mu$ expansion, $X^{(1)}$ corresponds to the Weyl-Ricci scalar (i.e.\ the $2d$ Weyl-Euler density) and $X^{(2)}$ corresponds to the classic ``$a=c$" result. For the Weyl anomaly in $6d$ and $8d$ both $X^{(3)}$ and $X^{(4)}$ can also be written as linear combinations of the Weyl-Euler density and type B anomalies. This is true for both the FG and WFG cases, just the quantities in the latter are Weyl-covariant. One just needs to substitute the Weyl quantities with their LC counterparts (i.e.\ set $a_\mu$ to zero) to get the Weyl anomaly in the FG case. However, when expressing them in terms of the Weyl-Schouten tensor and extended Weyl-obstruction tensors (or Schouten tensor and extended obstruction tensors in the FG case), we observe that the polynomial terms of $X^{(k)}/L^{2k}$ (without the total derivative terms) in $2k$-dimensions, denoted by $\bar X^{(k)}$, have the following structures:
\begin{align}
\label{X1D}
\bar X^{(1)}&=-\delta^\mu_\nu\hat P^\nu{}_\mu\,,\\
2\bar X^{(2)}&=\frac{1}{2}\delta^{\mu_1\mu_2}_{\nu_1\nu_2}\hat P^{\nu_1}{}_{\mu_1}\hat P^{\nu_2}{}_{\mu_2}\,,\\
6\bar X^{(3)}&=-\frac{1}{4}\delta^{\mu_1\mu_2\mu_3}_{\nu_1\nu_2\nu_3}\hat P^{\nu_1}{}_{\mu_1} \hat P^{\nu_2}{}_{\mu_2}\hat P^{\nu_3}{}_{\mu_3}-\frac{1}{2}\delta^{\mu_1\mu_2}_{\nu_1\nu_2}\hat\Omega_{(1)}^{\nu_1}{}_{\mu_1}\hat P^{\nu_2}{}_{\mu_2}\,,\\
\label{X4D}
24\bar X^{(4)}&=\frac{1}{8}\delta^{\mu_1\mu_2\mu_3\mu_4}_{\nu_1\nu_2\nu_3\nu_4}\hat P^{\nu_1}{}_{\mu_1} \hat P^{\nu_2}{}_{\mu_2}\hat P^{\nu_3}{}_{\mu_3}\hat P^{\nu_4}{}_{\mu_4}+\frac{1}{2}\delta^{\mu_1\mu_2\mu_3}_{\nu_1\nu_2\nu_3}\hat\Omega_{(1)}^{\nu_1}{}_{\mu_1}\hat P^{\nu_2}{}_{\mu_2}\hat P^{\nu_3}{}_{\mu_3}\nn\\
&\quad+\frac{1}{4}\delta^{\mu_1\mu_2}_{\nu_1\nu_2}\hat\Omega_{(1)}^{\nu_1}{}_{\mu_1}\hat\Omega_{(1)}^{\nu_2}{}_{\mu_2}+\frac{1}{4}\delta^{\mu_1\mu_2}_{\nu_1\nu_2}\hat\Omega_{(2)}^{\nu_1}{}_{\mu_1}\hat P^{\nu_2}{}_{\mu_2}\,,
\end{align}
where the Kronecker $\delta$ symbol is defined as 
\begin{align}
\delta^{\mu_1\cdots\mu_s}_{\nu_1\cdots\nu_s}=s!\delta^{\mu_1}{}_{[\nu_1}\cdots\delta^{\mu_s}{}_{\nu_s]}\,.
\end{align}
From \eqref{X1D}--\eqref{X4D} we can see that $\bar X^{(k)}$ contains all kinds of possible combinations of $\hat P_{\mu\nu}$ and $\hat\Omega^{(2<j<2k)}_{\mu\nu}$ whose Weyl weights add up to be $2k$, i.e.\ the Weyl weight of $X^{(k)}$. Using this pattern, one can directly write down the terms in the holographic Weyl anomaly in any dimension. For instance, we can easily predict without explicit calculation that $\bar X^{(5)}$ is the linear combination of the following terms:
\begin{align*}
&\delta^{\mu_1\mu_2\mu_3\mu_4\mu_5}_{\nu_1\nu_2\nu_3\nu_4\nu_5}\hat P^{\nu_1}{}_{\mu_1} \hat P^{\nu_2}{}_{\mu_2}\hat P^{\nu_3}{}_{\mu_3}\hat P^{\nu_4}{}_{\mu_4}\hat P^{\nu_5}{}_{\mu_5}\,,\quad\delta^{\mu_1\mu_2\mu_3\mu_4}_{\nu_1\nu_2\nu_3\nu_4}\hat\Omega_{(1)}^{\nu_1}{}_{\mu_1} \hat P^{\nu_2}{}_{\mu_2}\hat P^{\nu_3}{}_{\mu_3}\hat P^{\nu_4}{}_{\mu_4}\,,\\
&\delta^{\mu_1\mu_2\mu_3}_{\nu_1\nu_2\nu_3}\hat\Omega_{(2)}^{\nu_1}{}_{\mu_1}\hat P^{\nu_2}{}_{\mu_2}\hat P^{\nu_3}{}_{\mu_3}\,,\quad\delta^{\mu_1\mu_2\mu_3}_{\nu_1\nu_2\nu_3}\hat\Omega_{(1)}^{\nu_1}{}_{\mu_1}\hat\Omega_{(1)}^{\nu_2}{}_{\mu_2}\hat P^{\nu_3}{}_{\mu_3}\,,\quad\delta^{\mu_1\mu_2}_{\nu_1\nu_2}\hat\Omega_{(2)}^{\nu_1}{}_{\mu_1}\hat\Omega_{(1)}^{\nu_2}{}_{\mu_2}\,,\quad\delta^{\mu_1\mu_2}_{\nu_1\nu_2}\hat\Omega_{(3)}^{\nu_1}{}_{\mu_1}\hat P^{\nu_2}{}_{\mu_2}\,.
\end{align*}
These terms represent the independent central charges that appear in the holographic Weyl anomaly in $d=10$. 
\par
Based on the above pattern, it is natural to expect a general expression that can generate the holographic Weyl anomaly in any dimension, which is an analog of the exponential structure given by the Chern class that generates the chiral anomaly in any dimension (see, e.g.\ \cite{Frampton:1983nr,Zumino:1983rz,Bertlmann:1996xk}). It has been suggested in \cite{Deser1993} that the type A Weyl anomaly can be generated by a mechanism similar to that for the chiral anomaly. The expressions for the Weyl anomaly in terms of the (Weyl-) Schouten tensor and the extended (Weyl-) obstruction tensors suggest a similar mechanism for the holographic Weyl anomaly.

\section{The Role of Weyl Structure}
\label{Sec6}
Now that we have obtained the Weyl-obstruction tensors and Weyl anomaly, let us provide some observations on how the $a_{\mu}$ mode \eqref{aex} is involved. We have already mentioned that according to the FG theorem, this mode is pure gauge in the bulk. Now we have a few clear manifestations of this from our calculations. 
\par
The first one is that the subleading terms $a^{(2k)}_{\mu}$ with $k>0$ in the expansion of $a_\mu$ cannot be determined from the Einstein equations when $a^{(0)}_{\mu}$ is given. This is different from the expansion of $h_{\mu\nu}$ where the subleading terms $\gamma_{\mu\nu}^{(2k)}$ can be solved (on-shell) in terms of $\gamma_{\mu\nu}^{(0)}$.
\par
The second one is that $a_{\mu}$ appears only inside total derivatives in $X^{(k)}$, and thus represents cohomologically trivial modifications of the boundary Weyl anomaly. For $a_{\mu}^{(2k)}$ with $k\geqslant 2$, this can easily be seen from the expressions \eqref{4dAP}, \eqref{6dA} and \eqref{8dA}. What is not explicit in these formulas is that $a^{(0)}_{\mu}$ also appears inside a total derivative. This can be verified by separating the LC quantities out of the Weyl quantities in $X^{(k)}$. For instance, denote the LC Schouten tensor as $\mathring P_{\mu\nu}$ and the LC connection as $\LCnabla$, and then $X^{(1)}$ in $2d$ and $X^{(2)}$ in $4d$ can be written as
\begin{align}
L^{-2}X^{(1)}_{d=2}={}& L^{-2}\mathring X^{(1)}_{d=2}+\LCnabla\cdot a^{(0)}\,,\\
L^{-4}X^{(2)}_{d=4}={}&L^{-4}\mathring X^{(2)}_{d=4}-\frac{1}{2}\LCnabla_\mu(\mathring P^{\mu\nu}a^{(0)}_\nu-\mathring Pa_{(0)}^\mu)\nn\\
&-\frac{1}{4}\LCnabla_\mu(a^{(0)}_\nu\LCnabla^\nu a_{(0)}^\mu-a_{(0)}^\mu\LCnabla\cdot a_{(0)})-\frac{1}{4}\LCnabla_\mu(a_{(0)}^\mu a_{(0)}^2)-\frac{1}{2L^2}\LCnabla\cdot a^{(2)}\,,
\end{align}
where $L^{-2}\mathring X^{(1)}=-\mathring{P}$ and $L^{-4}\mathring X^{(2)}=\frac{1}{4}\mathring{P}^2-\frac{1}{4}\tr(\mathring{P}^2)$.\footnote{Note that $\LCnabla\cdot a^{(2)}$ is equivalent to $\hat\nabla\cdot a^{(2)}$ in $4d$, since in $2k$-dimension $\hat\nabla$ and $\LCnabla$ give the same result when acting on a vector with Weyl weight $+2k$ (see Appendix \ref{WG}).}
Notice that although the terms involving $a^{(0)}_{\mu}$ are total derivatives, they are not Weyl-covariant and so one cannot naively assume that they are trivial cocycles. However, by finding suitable local counterterms, we have checked that all the terms involving $a_{\mu}^{(0)}$ are indeed part of a trivial cocycle  for $2d$ and $4d$. As $a_{\mu}$ is pure gauge, we expect this to be generally true.
\par
In principle, the Weyl connection $a^{(0)}_\mu$ on the boundary brings new Weyl-invariant objects, such as $\tr(f_{(0)}^2)$, which could lead to new central charges in the Weyl anomaly. However, up to $d=8$ we find the classification of type A and type B anomalies is still available, and in such a basis the nonvanishing central charges are still the same as those in the FG case. Once this can be carried over to higher dimensions, then $a^{(0)}_\mu$ appearing in total derivatives in $X^{(k)}$ can also be deduced by considering the Weyl anomaly as the sum of the type A and type B anomalies. In the FG gauge, under a Weyl transformation the type B anomaly is invariant while the type A anomaly, i.e.\ the Euler density, gets an extra total derivative involving $\ln{\cal B}$. Since the Weyl connection makes the Weyl anomaly in the WFG gauge Weyl-invariant, the terms with $a^{(0)}_\mu$ in the Weyl-Euler density should exactly compensate the extra total derivative, and hence they must form a total derivative. 
\par
Another observation we have mentioned is that although the subleading terms in the expansion of $a_{\mu}$ make an appearance in $\gamma^{(2k)}_{\mu\nu}$, they do not appear in the Weyl-obstruction tensors. Up to $k=3$, we have seen explicitly in \eqref{g2}, \eqref{g4} and \eqref{g6} that the terms with $a_{\mu}^{(2)}$ and $a_{\mu}^{(4)}$ do not contribute to the pole at $d=2k$ in $\gamma^{(2k)}_{\mu\nu}$. What is also true but not as obvious, is that the terms with $a^{(0)}_\mu$ do not contribute to the pole at $d-2$ in the Weyl-Schouten tensor and are proportional to $d-2k$ in Weyl-obstruction tensors. For instance, one can separate the $a^{(0)}_\mu$ from $\hat P_{\mu\nu}$ and get
\begin{align}
\hat P_{\mu\nu}=\mathring P_{\mu\nu}+\LCnabla_\nu a^{(0)}_{\mu}+a^{(0)}_{\mu}a^{(0)}_{\nu}-\frac{1}{2}a_{(0)}^2\gamma^{(0)}_{\mu\nu}\,,
\end{align}
while the only pole on the right-hand side is in the LC Schouten tensor $\mathring P_{\mu\nu}$. Similarly, expressing the Weyl-Bach tensor in terms of LC quantities we have
\begin{align}
\hat B_{\mu\nu}=\mathring{B}_{\mu\nu}+(d-4)(a_{(0)}^\lambda\mathring{C}_{\lambda\nu\mu}-2a_{(0)}^\lambda\mathring{C}_{\mu\nu\lambda}+a_{(0)}^\lambda a_{(0)}^\rho\mathring W_{\rho\mu\lambda\nu})\,.
\end{align}
Thus, when $d=4$, $a^{(0)}_\mu$ does not contribute to the pole in $\gamma^{(4)}_{\mu\nu}$, and the Weyl-Bach tensor $\hat{B}_{\mu\nu}$ is equivalent to the LC Bach tensor $\mathring{B}_{\mu\nu}$. One should naturally expect that this is also true for any Weyl-obstruction tensors, i.e.\ $\hat{\cal O}^{(2k)}_{\mu\nu}$ is equivalent to the LC obstruction tensor $\mathring{\cal O}^{(2k)}_{\mu\nu}$ when $d=2k$. Note that when $d>2k$, the $a^{(0)}_\mu$ terms are included in the Weyl-obstruction tensor so that $\hat{\cal O}^{(2k)}_{\mu\nu}$ is always Weyl-covariant.
\par
The statement that any term in the expansion of $a_\mu$ does not appear in the pole of $\hat\gamma^{(2k)}_{\mu\nu}$ is consistent with the following claim: when $d=2k$, the Weyl-obstruction tensor $\hat{\cal O}_{(2k)}^{\mu\nu}$ satisfies
\begin{align}
\label{varX}
\hat{\cal O}_{(2k)}^{\mu\nu}=\frac{1}{\sqrt{-\det\gamma^{(0)}}}\frac{\delta}{\delta\gamma^{(0)}_{\mu\nu}}\int\td^d x\sqrt{-\det\gamma^{(0)}}X^{(k)}\,.
\end{align}
The FG version of this relation for $\mathring{\cal O}_{(2k)}^{\mu\nu}$ was proved in \cite{graham2005ambient} (see also \cite{deHaro:2000vlm}). If the claim above can be proved for the WFG gauge, then the reason that none of the terms in the expansion of $a_\mu$ contributes to $\hat{\cal O}^{(2k)}_{\mu\nu}$ at $d=2k$ will be straightforward: as they only appear in  total derivative terms in $X^{(k)}$, they will be dropped in the variation above. Hence, this can be viewed as another manifestation of $a_\mu$ being pure gauge in the bulk. We have verified by brute force that for $k=2$ the variation in \eqref{varX} indeed gives the Weyl-Bach tensor when $d=4$, and a rigorous proof for any $k$ is worth further study. 
\par
Based on the FG version of relation \eqref{varX}, there is another approach of finding the (LC) obstruction tensors and Weyl anomaly in even dimensions called the dilatation operator method \cite{Anastasiou:2020zwc}. As a consistency check, we also computed the $8d$ Weyl anomaly in the FG gauge using this method. We will briefly introduce this method in Appendix \ref{AppC} and show there that the result in $8d$ agrees with what we have in \eqref{8dA} when the Weyl structure is turned off.

\section{Conclusions}
\label{Sec7}
In this work, we first derived the obstruction tensor from the pole at $d=2k$ of the (on-shell) $\gamma^{(2k)}_{\mu\nu}$ in the FG expansion of an AlAdS spacetime using dimensional regularization. Under an appropriate analytical continuation when $d$ approaches an even integer, this approach is equivalent to the one with a logarithmic term in \cite{Fefferman2011,graham2005ambient}. We defined the pole term in the expansion to be Graham's extended obstruction tensor, whose residue is the obstruction tensor (up to a constant factor). Then, after introducing the WFG ansatz, we generalized the Schouten tensor and obstruction tensors in the FG gauge to the Weyl-Schouten tensor and Weyl-obstruction tensors in the WFG gauge, which are now Weyl-covariant in any dimension. By solving the bulk Einstein equations, we computed the Weyl-obstruction tensors in $4d$ (i.e.\ the Weyl-Bach tensor) and $6d$ explicitly, and found that they have almost the same form as the corresponding obstruction tensors, with everything Weyl-covariantized and some extra terms due to the Weyl-Schouten tensor being not symmetric. This is a natural manifestation of the fact that the WFG gauge Weyl-covariantizes the boundary geometry. We  observed that all the subleading terms in the expansion of $a_{\mu}$ do not contribute to the Weyl-obstruction tensor. We also found that when $d=2k$, the Weyl-obstruction tensor ${\cal O}^{(2k)}_{\mu\nu}$ is equivalent to its LC counterpart, and so $a^{(0)}_{\mu}$ does not contribute to the obstruction either. When $d>2k$, the $a^{(0)}_{\mu}$ terms are included in the ${\cal O}^{(2k)}_{\mu\nu}$ to make it Weyl-covariant.
\par
As the main result of this paper, we computed the Weyl anomaly in $6d$ and $8d$ in the WFG gauge by using the Weyl-Schouten tensor and extended Weyl-obstruction tensors as the building blocks. The Weyl anomaly shown in \eqref{6dA} and \eqref{8dA} indeed go back to the corresponding FG results when the Weyl structure $a_\mu$ is turned off, but now they become Weyl-covariant. In addition, we also re-expressed the Weyl anomaly in $2d$ and $4d$ in terms of the Weyl-Schouten tensor. By observing the pattern of the Weyl anomaly in different dimensions, we suspect there exists a general formulation that can generate the holographic Weyl anomaly in any dimension, which will be explored in future work.
\par
In the boundary field theory, both the induced metric $\gamma^{(0)}_{\mu\nu}$ and the Weyl connection $a^{(0)}_\mu$ are non-dynamical background fields. However, only $\gamma^{(0)}_{\mu\nu}$ is sourcing a current operator, namely the energy-momentum tensor, while $a^{(0)}_\mu$ does not source any current since $a_\mu$ is pure gauge in the bulk. From the Weyl-Ward identity \eqref{holoWeylWard}, we can see that the trace of the energy-momentum tensor obtains a contribution from $p^{(0)}_\mu$ due to the gauge freedom of WFG. Together we can regard it as an improved energy-momentum tensor $\tilde T_{\mu\nu}$. For non-holographic field theories with background Weyl geometry the corresponding Weyl current $J^\mu$ of the Weyl connection does not need to vanish. The Weyl current in the general case deserves further investigation.
\par
An important corollary in our analysis is that the Weyl structure $a_{\mu}$ only appears as a trivial cocycle in the Weyl anomaly, and thus only contributes  cohomologically trivial modifications. From the Weyl anomaly up to $8d$ we can directly see this for the subleading terms of $a_{\mu}$ as they appear only in total derivative terms in $X^{(k)}$. For the leading term $a_{\mu}^{(0)}$ this is less obvious since it plays the role of the boundary Weyl connection, but one can verify that by writing the anomaly in terms of the boundary LC connection, the terms involving $a_{\mu}^{(0)}$ also represent trivial cocycles. This indicates a striking feature of the WFG gauge, namely $a^{(0)}_\mu$ manages to make the expressions Weyl-covariant without introducing new central charges, which, once again, is consistent with the fact that $a_\mu$ is pure gauge in the bulk. Nonetheless, these cohomologically trivial terms might have significant effects in the presence of corners, i.e.\ spacelike codimension-2 surfaces.\footnote{We thank Rob Leigh and Luca Ciambelli for pointing this out in conversations.} The recent construction proposed in \cite{Ciambelli:2021vnn,Freidel:2021cbc} may be useful for the analysis of these effects.
\par
In this paper we introduced the obstruction tensor and extended obstruction tensor as the pole of $\gamma^{(2k)}_{\mu\nu}$. However, as we have mentioned, they can also be defined using the ambient construction. What we have found but not demonstrated in this paper is that the Weyl-obstruction tensors and extended Weyl-obstruction tensors can be defined in a similar way by promoting the ambient metric to the Weyl-ambient metric. We expect to discuss the Weyl-ambient construction in detail in a future publication.
\par
Finally, although this paper focuses on the holographic Weyl anomaly, we believe that the  (Weyl-) Schouten tensor and extended (Weyl-) obstruction tensors can also be used as the building blocks for the Weyl anomaly of other theories in general. How can these building blocks arise in a non-holographic context requires a deep understanding of the Lorentz-Weyl structure of a frame bundle, which encodes all the local Lorentz and Weyl transformations. To achieve this, the picture of Atiyah Lie algebroids introduced in \cite{Ciambelli:2021ujl} for gauge theories can be used to organize the Weyl and Lorentz anomalies in a geometric fashion. By means of this geometric picture, we look forward to carrying over the holographic results obtained in this paper to the construction of Weyl anomaly in the general case.

\section*{Acknowledgements}
We would like to thank Rob Leigh for suggesting the problem and providing constant support to us. We are also grateful to Luca Ciambelli for many valuable discussions and carefully going through our manuscript. This work was partially supported by the U.S. Department of Energy under contract DE-SC0015655.

\begin{appendix}
\section{Weyl Geometry}
\label{WG}
This appendix provides a brief review of Weyl geometry\cite{Folland:1970,Hall:1992}. We will mainly introduce the geometric quantities equipped with Weyl connection as well as some useful relations we used in the previous sections. We use $a,b,\cdots$ to label the internal Lorentz indices and $\mu,\nu,\cdots$ to label the spacetime indices. For clarity, we also put $\circ$ on the top of LC quantities, e.g.\ $\mathring R^a{}_{bcd}$, $\mathring P_{ab}$, etc.
\par
Given a generalized Riemannian manifold $(M,g)$ with a connection $\nabla$, in an arbitrary basis $\{\un e_a\}$, the connection coefficients $\Gamma^c{}_{ab}$ are defined as
\begin{align}
\label{conncoef}
\nabla_{\un e_a}\un e_b=\Gamma^c{}_{ab}\un e_c\,.
\end{align}
The torsion tensor and Riemann curvature tensor of $\nabla$ in this basis are given by
\begin{align}
\label{Tor}
T^c{}_{ab}\un e_c&\equiv \nabla_{\un e_a}\un e_b-\nabla_{\un e_b}\un e_a-[\un e_a,\un e_b]\,,\\
\label{Rie}
R^a{}_{bcd}\un e_a&\equiv \nabla_{\un e_c}\nabla_{\un e_d}\un e_b-\nabla_{\un e_d}\nabla_{\un e_c}\un e_b-\nabla_{[\un e_c,\un e_d]}\un e_b\,.
\end{align}
When $\nabla$ is associated with $g$ and is torsion-free, it is called a Levi-Civita (LC) connection, denoted by $\LCnabla$. Using $\LCGamma$ to denote the LC connection coefficients, we have $\mathring\nabla_{\un e_a}\un e_b=\LCGamma^c{}_{ab}\un e_c$. 
By definition, the conditions satisfied by the LC connection coefficients $\LCGamma^c{}_{ab}$ are
\begin{align}
\label{NM}
0&=(\LCnabla g)(\un e_a,\un e_b,\un e_c)=\LCnabla_{\un e_c}g(\un e_a,\un e_b)-\LCGamma^d{}_{ca}g(\un e_d,\un e_b)-\LCGamma^d{}_{cb}g(\un e_d,\un e_a)\,,\\
0&=T^a{}_{bc}=\LCGamma^c{}_{ab}-\LCGamma^c{}_{ba}-C_{ab}{}^c\,,
\end{align}
where $C_{\mu\nu}{}^\rho$ are the commutation coefficients defined by $[\un e_a,\un e_b]=C_{ab}{}^c\un e_c$. Denote $g_{ab}\equiv g(\un e_a,\un e_b)$ as the component of the metric in the frame $\{\un e_a\}$. From these conditions $\LCGamma^c{}_{ab}$ can be derived as
\begin{align}
\label{LC}
\LCGamma^c{}_{ab}=&\,\frac{1}{2}g^{cd}\big(\un e_a(g_{db})+\un e_b(g_{ad})-\un e_d(g_
{ab})\big)-\frac{1}{2}g^{cd}(C_{ad}{}^e g_{eb}+C_{bd}{}^e g_{ae}-C_{ab}{}^e g_{ed})\,.
\end{align}
\par
Now we will work in a coordinate basis $\{\un\p_\mu\}$.\footnote{Note that $\un e_a\equiv e_a^\mu\un\p_\mu$ and $\bm e^a\equiv e^a_\mu\td x^\mu$ have Weyl weights $+1$ and $-1$ respectively, while $\un\p_\mu$ and $\td x^\mu$ have no Weyl weights. This is because the Weyl transformation of the frame only comes from the soldering of the vector bundle associated with the frame bundle to the tangent space of $M$.} Consider a Weyl transformation
\begin{align}
\label{WeylA}
g\to{\cal B}^{-2}g\,.
\end{align}
The metricity tensor $\nabla g$ will not transform covariantly under \eqref{WeylA}. To restore the Weyl covariance, one can introduce a Weyl connection $A=A_\mu\td x^\mu$ which transforms under a Weyl transformation as
\begin{align}
A_\mu\to A_\mu-\nabla_\mu\ln{\cal B}\,.
\end{align}
Then, we obtain an object that is Weyl-covariant:
\begin{align}
(\nabla_\mu g_{\nu\rho}-2A_\mu g_{\nu\rho})\to{\cal B}^{-2}(\nabla_\mu g_{\nu\rho}-2A_\mu g_{\nu\rho})\,.
\end{align}
More generally, for a tensor $T$ of an arbitrary type (with indices suppressed) that transforms under a Weyl transformation with a specific Weyl weight $\omega_T$, i.e.\ $T\to B^{\omega_T}T$, we can define
\begin{align}
\label{WeylD}
\hat\nabla_\mu T\equiv\nabla_\mu T+w_TA_\mu T\,.
\end{align}
In this way, $\hat\nabla$ acting on $T$ will also transform Weyl-covariantly as $\hat\nabla_\mu T\to B^{\omega_T}\hat\nabla_\mu T$.
\par
Now we choose the connection $\nabla$ by setting the Weyl metricity as follows
\begin{align}
0&=\nabla_\mu g_{\nu\rho}-2A_\mu g_{\nu\rho}=\hat\nabla_\mu g_{\nu\rho}\,.
\end{align}
We will also require $\nabla$ defined in the above equation to be torsion-free. With the existence of the Weyl metricity, the connection coefficients of $\nabla$ in the coordinate basis become
\begin{align}
\label{WeylLC}
\Gamma^\rho{}_{\mu\nu}={}&\frac{1}{2}g^{\rho\sigma}(\p_\mu g_{\sigma\nu}+\p_\nu g_{\nu\sigma}-\p_\sigma g_
{\mu\nu})-(A_\mu\delta^\rho{}_\nu+A_\nu\delta^\rho{}_\mu-g^{\rho\sigma}A_\sigma g_{\mu\nu})\,.
\end{align}
We can see that this is different from the familiar Christoffel symbols due to the extra terms involving the Weyl connection. When $\nabla$ and $\LCnabla$ act on a vector, their difference can be reflected by
\begin{align}
\label{div_v}
\nabla_\mu v^\nu=\LCnabla_\mu v^\nu-(A_\mu\delta^\nu{}_\rho+A_\rho\delta^\nu{}_\mu-g^{\nu\sigma}A_\sigma g_{\mu\rho})v^\rho\,.
\end{align}
It is worthwhile to notice that if $v^\nu$ has Weyl weight $d=\dim M$, then it follows from \eqref{WeylD} and \eqref{div_v} that $\hat\nabla_\mu v^\mu=\LCnabla_\mu v^\mu$.
\par
Now one can compute the Riemann tensor of $\nabla$ and its contractions. Denoting the coordinate components of the Riemann tensor of $\LCnabla$ as $\mathring R^\mu{}_{\nu\rho\sigma}$, one finds from \eqref{Rie} that
\begin{align}
\label{WRiem}
R^\mu{}_{\nu\rho\sigma}={}&\mathring{R}^\mu{}_{\nu\rho\sigma}+\LCnabla_\sigma A_\nu\delta^\mu{}_\rho-\LCnabla_\rho A_\nu\delta^\mu{}_\sigma
+(\LCnabla_\sigma A_\rho-\LCnabla_\rho A_\sigma)\delta^\mu{}_\nu
+\LCnabla_\rho A^\mu g_{\nu\sigma}-\LCnabla_\sigma A^\mu g_{\nu\rho}\nn\\
&+A_\nu(A_\sigma\delta^\mu{}_\rho-A_\rho\delta^\mu{}_\sigma)
+A^\mu(g_{\nu\sigma}A_\rho-g_{\nu\rho}A_\sigma)
+A^2(g_{\nu\rho}\delta^\mu{}_\sigma-g_{\nu\sigma}\delta^\mu{}_\rho)\,,\\
R_{\mu\nu}={}&\mathring{R}_{\mu\nu}-\frac{d}{2}F_{\mu\nu}+(d-2)(\LCnabla_{(\mu}A_{\nu)}+A_\mu A_\nu)+(\LCnabla\cdot A-(d-2)A^2)g_{\mu\nu}\,,\\
\label{WRic}
R={}&\mathring{R}+2(d-1)\LCnabla\cdot A-(d-1)(d-2)A^2\,,
\end{align}
where $R_{\mu\nu}\equiv R^\rho{}_{\mu\rho\nu}$, $R\equiv R_{\mu\nu}g^{\mu\nu}$, and we defined the curvature of $A_\mu$ as $F_{\mu\nu}=\LCnabla_{\mu}A_{\nu}-\LCnabla_{\nu}A_{\mu}$. It is easy to see from \eqref{WRiem} that, unlike $\mathring R^\mu{}_{\nu\rho\sigma}$, the $R^\mu{}_{\nu\rho\sigma}$ of $\nabla$ now is not antisymmetric in the first two indices, and it does not have the interchange symmetry for the two index pairs. Also, the $R_{\mu\nu}$ of $\nabla$ is not symmetric due to the appearance of the $F_{\mu\nu}$ term.
\par
On the other hand, from \eqref{conncoef} we have the connection coefficients $\hat\Gamma^c{}_{ab}$ for $\hat\nabla$:
\begin{align}
\hat\Gamma^c{}_{ab}\un e_c=\hat\nabla_{\un e_a}\un e_b=\nabla_{\un e_a}\un e_b+A(\un e_a)\un e_b=\Gamma^c{}_{ab}\un e_c+A(\un e_a)\un e_b\,,
\end{align}
where we used the fact that the basis vector $\un e_a$ has Weyl weight $+1$. Plugging this into \eqref{Rie}, we find that the Riemann tensor of $\hat\nabla$ and its contractions satisfy
\begin{align}
\label{hatR}
\hat R^\mu{}_{\nu\rho\sigma}=&R^\mu{}_{\nu\rho\sigma}+\delta^\mu{}_\nu F_{\rho\sigma}\,,\qquad\hat R_{\mu\nu}=R_{\mu\nu}+F_{\mu\nu}\,,\qquad\hat R=R\,.
\end{align}
We refer to $\hat R^\mu{}_{\nu\rho\sigma}$, $\hat R_{\mu\nu}$ and $\hat R$ as the Weyl-Riemann tensor, Weyl-Ricci tensor, and Weyl-Ricci scalar, respectively.\footnote{Note that this is different from \cite{Ciambelli:2019bzz}, in which the quantities defined using $\nabla$ instead of $\hat\nabla$ are called Weyl quantities.} Similar to the curvature tensors for $\nabla$, the Weyl-Riemann tensor is not antisymmetric in the first two indices and does not have the interchange symmetry for the two index pairs, and the Weyl-Ricci tensor is not symmetric. Also notice that the Weyl-Weyl tensor, namely the traceless part of the Weyl-Riemann tensor, is equal to the LC Weyl tensor, i.e.
\begin{align}
\hat W^\mu{}_{\nu\rho\sigma}=\mathring W^\mu{}_{\nu\rho\sigma}\,.
\end{align}
\par
Unlike the LC curvature quantities, which transform in a non-covariant way under the Weyl transformation, the Weyl-Riemann tensor, Weyl-Ricci tensor, and Weyl-Ricci scalar transform under the Weyl transformation as
\begin{align}
\label{hatRtrans}
\hat R^\mu{}_{\nu\rho\sigma}\to\hat R^\mu{}_{\nu\rho\sigma}\,,\qquad\hat R_{\mu\nu}\to\hat R_{\mu\nu}\,,\qquad\hat R\to{\cal B}^2\hat R\,.
\end{align}
Furthermore, we can define the Weyl-Schouten tensor $\hat P_{\mu\nu}$ and Weyl-Cotton tensor $\hat C_{\mu\nu\rho}$ as
\begin{align}
\label{WP1}
\hat P_{\mu\nu}&=\frac{1}{d-2}\bigg(\hat R_{\mu\nu}-\frac{1}{2(d-1)}\hat Rg_{\mu\nu}\bigg)\,,\\
\label{WC1}
\hat C_{\mu\nu\rho}&=\hat\nabla_{\rho}\hat P_{\mu\nu}-\hat\nabla_{\nu}\hat P_{\mu\rho}\,.
\end{align}
Although the LC Schouten tensor $\mathring P_{\mu\nu}$ defined by substituting $\hat R_{\mu\nu}$ and $\hat R$ in \eqref{WP1} with $R_{\mu\nu}$ and $R$ is a symmetric tensor, $\hat P_{\mu\nu}$ has an antisymmetric part $\hat P_{[\mu\nu]}=-F_{\mu\nu}/2$. In terms of the LC connection, the Bach tensor is defined by (the indices of the components are raised and lowered by $g$)
\begin{align}
\mathring B_{\mu\nu}=\LCnabla^\rho\LCnabla_\rho \mathring P_{\mu\nu}-\LCnabla^\rho\LCnabla_{\nu}\mathring P_{\mu\rho}-\mathring W_{\sigma\nu\mu\rho}\mathring P^{\rho\sigma}\,,
\end{align}
which satisfies $\mathring B_{\mu\nu}\to{\cal B}^2\mathring B_{\mu\nu}$ in $4d$. Now we can define the Weyl-Bach tensor
\begin{align}
\label{WB1}
\hat B_{\mu\nu}=\hat\nabla^\rho\hat\nabla_\rho \hat P_{\mu\nu}-\hat\nabla^\rho\hat\nabla_{\nu}\hat P_{\mu\rho}-\hat W_{\sigma\nu\mu\rho}\hat P^{\rho\sigma}\,.
\end{align}
Similar to the LC Bach tensor, the Weyl-Bach tensor is also symmetric and traceless; however, it is Weyl-covariant in any dimension. Following \eqref{WRiem}--\eqref{WRic}, here we list the above-mentioned Weyl quantities in terms of their corresponding LC quantities:
\begin{align}
\label{hatP}
\hat P_{\mu\nu}&=\mathring P_{\mu\nu}+\LCnabla_\nu A_{\mu}+A_{\mu}A_{\nu}-\frac{1}{2}A^2g_{\mu\nu}\,,\\
\hat C_{\mu\nu\rho}&=\mathring C_{\mu\nu\rho}-A_\sigma\mathring W^\sigma{}_{\mu\rho\nu}\,,\\
\hat B_{\mu\nu}&=\mathring{B}_{\mu\nu}+(d-4)(A^\rho\mathring{C}_{\rho\nu\mu}-2A^\rho\mathring{C}_{\mu\nu\rho}+A^\rho A^\sigma \mathring W_{\sigma\mu\rho\nu})\,.
\end{align}
\par
The Bianchi identity for $\hat\nabla$ reads
\begin{align}
\hat\nabla_\mu \hat R^\lambda{}_{\nu\rho\sigma}+\hat\nabla_\rho\hat R^\lambda{}_{\nu\sigma\mu}+\hat\nabla_\sigma\hat R^\lambda{}_{\nu\mu\rho}=0\,.
\end{align}
Noticing that $\hat\nabla_\mu g_{\nu\rho}=0$, the contraction of the above equation gives
\begin{align}
\label{BI}
\hat\nabla^\mu\hat G_{\mu\nu}=0\,,
\end{align}
where we defined the Weyl-Einstein tensor $\hat G_{\mu\nu}\equiv \hat R_{\mu\nu}-\frac{1}{2}\hat Rg_{\mu\nu}$. Using \eqref{WP1}, this identity can also be expressed using the Weyl-Schouten tensor as
\begin{align}
\label{BIP}
\hat\nabla^\mu\hat P_{\mu\nu}=\hat\nabla_\nu\hat P\,.
\end{align}
where $\hat P$ is the trace of $\hat P_{\mu\nu}$. Starting from \eqref{WB1} and using \eqref{BIP} repeatedly, one obtains
\begin{align}
\label{nablaB}
\hat\nabla^\mu\hat B_{\mu\nu}=(d-4)\hat P^{\mu\rho}(\hat C_{\rho\mu\nu}+\hat C_{\nu\mu\rho})\,.
\end{align} 
Note that since  $\mathring P$ is symmetric, the above equation in the LC case becomes
\begin{align}
\LCnabla^\mu\mathring B_{\mu\nu}=(d-4)\mathring P^{\mu\rho}\mathring C_{\rho\mu\nu}\,.
\end{align} 
It is also useful to notice that in the LC case, the divergence of the Cotton tensor vanishes
\begin{align}
\label{divC}
\LCnabla^\mu\mathring C_{\mu\nu\rho}=0\,,
\end{align} 
while for the Weyl-Cotton tensor we have instead
\begin{align}
\hat\nabla^\mu\hat C_{\mu\nu\rho}=\hat W_{\sigma\rho\lambda\nu}F^{\sigma\lambda}\,.
\end{align} 
In the end of this appendix, we list the Weyl weights of the above-mentioned Weyl quantities:
\begin{table}[!htbp]
\centering
\caption{Weyl weights of Weyl-covariant quantities}
\begin{tabular}{ccccccccccc}
\toprule
$\un e_a$ & $\bm e^a$ & $g_{\mu\nu}$ & $g^{\mu\nu}$ &$\hat R^\mu{}_{\nu\rho\sigma}$ & $\hat R_{\mu\nu}$ & $\hat R$ & $F_{\mu\nu}$& $\hat P_{\mu\nu}$ &$\hat C_{\mu\nu\rho}$ &$\hat B_{\mu\nu}$\\
\midrule
$+1$ & $-1$ & $-2$ & $+2$ &$0$ & $0$&$+2$ & $0$& $0$ &$0$ &$+2$\\
\bottomrule
\end{tabular}
\label{table1}
\end{table}

\section{Solving the Bulk Einstein Equations}
\label{AppB0}
To solve for $\gamma^{(2k)}_{\mu\nu}$ in the WFG gauge from the Einstein equations, we first introduced the following notations:
\begin{align}
\varphi_\mu&\equiv D_za_\mu\,,\qquad f_{\mu\nu}\equiv D_\mu a_\nu-D_\nu a_\mu\,,\qquad\rho_{\mu\nu}\equiv\frac{1}{2}D_zh_{\mu\nu}\,,\qquad\theta\equiv\tr\rho\,,\nn\\
\label{quantities}
\psi_{\mu\nu}&\equiv\rho_{\mu\nu}+\frac{L}{2}f_{\mu\nu}\,,\qquad
\gamma^\lambda{}_{\mu\nu}\equiv\Gamma^\lambda{}_{\mu\nu}=\frac{1}{2}h^{\lambda\rho}( D_\mu h_{\rho\nu} +D_\nu h_{\mu\rho} - D_\rho h_{\nu\mu})\,.
\end{align}
Since the integral curves of $\un D_z$ form a congruence, some of these quantities can be interpreted as the properties of this congruence: $\varphi^\mu$ is the acceleration, $f_{\mu\nu}$ is the twist, $\theta$ is the expansion and $\sigma_{\mu\nu}\equiv\rho_{\mu\nu}-\frac{1}{d}\theta h_{\mu\nu}$ is the shear. By plugging in the expansions \eqref{hex} and \eqref{aex}, one can obtain the expansions of the quantities above. A list of these expansions enough for capturing the first two leading orders of the Einstein equations can be found in the Appendix of \cite{Ciambelli:2019bzz}.
\par
Using the connection coefficients $\Gamma^\lambda{}_{\mu\nu}$ in the bulk, one can compute the curvature tensors and the Einstein tensor. Then, the vacuum Einstein equations can be written as
\begin{align}
\label{eomzz}
0&=G_{zz}+g_{zz}\Lambda=-\frac{1}{2}\tr(\rho\rho)-\frac{3L^2}{8}\tr(ff)-\frac{1}{2}\bar{R}+\frac{1}{2}\theta^2+\Lambda\,\\
\label{eomzm}
0&=G_{z\mu}+g_{z\mu}\Lambda=\nabla_\nu\psi^\nu{}_\mu-D_\mu\theta+L^2f_{\nu\mu}\varphi^\nu\,,\\
\label{eommn}
0&=G_{\mu\nu}+g_{\mu\nu}\Lambda=\bar{G}_{\mu\nu}-(D_z+\theta)\psi_{\mu\nu}-L\nabla_\nu\varphi_\mu+2\rho_{\nu\rho}\rho^\rho{}_\mu+\frac{L^2}{2}f_{\nu\rho}f^\rho{}_\mu-L^2\varphi_\mu\varphi_\nu\nn\\
&\qquad\qquad\qquad\qquad+h_{\mu\nu}\left(L\nabla_\mu\varphi^\mu+D_z\theta+\frac{1}{2}\tr(\rho\rho)-\frac{L^2}{8}\tr(ff)+L^2\varphi^2+\frac{1}{2}\theta^2+\Lambda\right)\,.
\end{align}
where $\Lambda=-\frac{d(d-1)}{2L^2}$ is the cosmological constant, and $\bar{R}=h^{\mu\nu}\bar R_{\mu\nu}$ with
\begin{align}
\bar{R}_{\mu\nu}=D_\rho\gamma^\rho
{}_{\nu\mu}-D_\nu\gamma^\rho{}_{\rho\mu}+\gamma^{\rho}{}_{\rho\sigma}\gamma^{\sigma}{}_{\nu\mu}-\gamma^{\rho}{}_{\nu\sigma}\gamma^{\sigma}{}_{\rho\mu}\,.
\end{align}
Denote $m_{(2k)\nu}^{\mu}\equiv\gamma_{(0)}^{\mu\rho}\gamma^{(2k)}_{\rho\nu}$ and $n_{(2k)\nu}^{\mu}\equiv\gamma_{(0)}^{\mu\rho}\pi^{(2k)}_{\rho\nu}$. Expanding \eqref{eomzz}--\eqref{eommn} using \eqref{hex} and \eqref{aex}, one can solve the Einstein equations order by order. First, the $zz$-component of the Einstein equations gives
\begin{align}
0={}&\bigg[\frac{d(d-1)}{2L^2}+\Lambda\bigg]-\frac{z^2}{L^2}\bigg[\frac{R^{(0)}}{2}+\frac{d-1}{L^2}X^{(1)}\bigg]+\frac{z^4}{L^4}\bigg[\frac{d}{2L^2}(X^{(1)})^2-\frac{2(d-1)}{L^2}X^{(2)}-\frac{1}{2L^2}\tr(m_{(2)}^2)\nn\\
&-\frac{3L^2}{8}\tr(f_{(0)}^2)-\frac{1}{2}\Big(\gamma_{(0)}^{\lambda\nu}\hat\nabla^{(0)}_\lambda\hat\nabla_\mu \big(m_{(2)}{}^{\mu}{}_{\nu}-\tr (m_{(2)})\delta^\mu{}_\nu\big)
+2(d-1)\hat\nabla\cdot a^{(2)}
-\tr\big(m_{(2)}\gamma_{(0)}^{-1}R^{(0)}\big)\Big)\bigg]\nn\\
&+\cdots-\frac{z^d}{L^d}(d-1)\bigg[\frac{d}{2L^2}Y^{(1)}+\hat\nabla\cdot p_{(0)}\bigg]+\cdots\,,
\end{align}
where $X^{(1)}$, $X^{(2)}$ and $Y^{(1)}$ are given in expansion \eqref{sqrth}, which can be expressed in terms of the expansion of $h_{\mu\nu}$ as
\begin{align}
\label{XY}
X^{(1)}&=\tr(m_{(2)})\,,\quad X^{(2)}=\tr(m_{(4)})-\frac{1}{2}\tr(m_{(2)}^2)+\frac{1}{4}\left(\tr(m_{(2)})\right)^2\,,\,\cdots\,,Y^{(1)}=\tr(n_{(0)})\,,\,\cdots\,.
\end{align}
At the $O(1)$-order, the $zz$-equation is trivially satisfied, and at the $O(z^2)$-order, we can find that
\begin{align}
X^{(1)}=-\frac{L^2}{2(d-1)} R^{(0)}=-L^2\hat P\,.
\end{align}
Then, using the above result we can obtain from the $O(z^4)$-order that
\begin{align}
X^{(2)}&=-\frac{1}{4}\tr(m_{(2)}^2)+\frac{1}{4}(X^{(1)})^2-\frac{L^2}{2}\hat\nabla\cdot a^{(2)}-\frac{L^4}{16}\tr(f^{(0)}f^{(0)})\nn\\
&=-\frac{L^4}{4}\tr(\hat P^2)+\frac{L^4}{4}\hat P^2-\frac{L^2}{2}\hat\nabla\cdot a^{(2)}\,,
\end{align}
where we used \eqref{Pgf}. Also notice that the $O(z^d)$-order gives the Weyl-Ward identity
\begin{align}
0=\frac{d}{2L^2}Y^{(1)}+\hat\nabla\cdot p_{(0)}\,.
\end{align}
\par
Now we look at the $\mu\nu$-components of the Einstein equations:
\begin{align}
0={}&\bigg[G^{(0)}_{\mu\nu}+\frac{d}{2}f^{(0)}_{\mu\nu}-\frac{d-2}{L^2}X^{(1)}\gamma^{(0)}_{\mu\nu}+\frac{d-2}{L^2}\gamma^{(2)}_{\mu\nu}\bigg]+\frac{z^2}{L^2}\bigg[\frac{1}{2}\hat\nabla_\lambda\Big(\gamma_{(0)}^{\lambda\xi}\Big(\hat\nabla_\nu \gamma^{(2)}_{\xi\mu}+\hat\nabla_\mu \gamma^{(2)}_{\xi\nu}-\hat\nabla_\xi \gamma^{(2)}_{\mu\nu}\Big)\Big)\nn\\
&-\frac{1}{2}\gamma^{(0)}_{\mu\nu}\hat\nabla_\mu\hat\nabla_\nu \Big(\gamma_{(2)}^{\mu\nu}-X^{(1)}\gamma_{(0)}^{\mu\nu}\Big)-\frac{1}{2}\hat\nabla_{(\mu}\hat\nabla_{\nu)} X^{(1)}+(d-4)\big(\hat\nabla^{(0)}_{(\mu} a_{\nu)}^{(2)}-\gamma_{\mu\nu}^{(0)}\hat\nabla\cdot a^{(2)})\nn\\
\label{emn}
&+\frac{2(d-4)}{L^{2}}\gamma^{(4)}_{\mu\nu}+\frac{2}{L^{2}}m_{(2)}^\rho{}_\mu\gamma^{(2)}_{\rho\nu}+\frac{L^2}{2}f^{(0)}_{\nu\rho}f^{(0)}_{\sigma\mu}\gamma_{(0)}^{\sigma\rho}+\Big(\frac{1}{2}\tr(m_{(2)}\gamma_{(0)}^{-1}{R}^{(0)})-\frac{L^2}{8}\tr(f^{(0)}f^{(0)})\nn\\
&-\frac{2(d-4)}{L^2}X^{(2)}+\frac{d-3}{2L^2}(X^{(1)})^2+\frac{1}{2L^2}\tr(m_ {(2)}^2)\Big)\gamma^{(0)}_{\mu\nu}\bigg]+\cdots\,.
\end{align}
Note that $\gamma_{(2)}^{\mu\nu}\equiv(\gamma_{(0)}^{-1}\gamma^{(2)}\gamma_{(0)}^{-1})^{\mu\nu}$ is not the inverse of $\gamma^{(2)}_{\mu\nu}$. Plugging in the results we got from the $zz$-equation, we obtain from the first two leading orders of \eqref{emn} that 
\begin{align}
\gamma^{(2)}_{\mu\nu}={}&-\frac{L^2}{d-2}\bigg(R^{(0)}_{(\mu\nu)}-\frac{1}{2(d-1)}R^{(0)}\gamma^{(0)}_{\mu\nu}\bigg)\,,\\
\gamma^{(4)}_{\mu\nu}={}&-\frac{L^2}{4(d-4)}\bigg(2\hat\nabla_\lambda\hat\nabla_{(\mu} m_{(2)}{}^{\lambda}{}_{\nu)}
-\hat\nabla\cdot\hat\nabla \gamma^{(2)}_{\mu\nu}-\hat\nabla_{(\mu}\hat\nabla_{\nu)} X^{(1)}-\frac{1}{L^2}\gamma^{(0)}_{\mu\nu}\tr(m_{(2)}^2)+\frac{4}{L^2}m_{(2)}^\rho{}_\mu\gamma^{(2)}_{\rho\nu}\nn\\
&\qquad\qquad\qquad+L^2f^{(0)}_{\nu\rho}f^{(0)}_{\sigma\mu}\gamma_{(0)}^{\sigma\rho}-\frac{L^2}{4}\tr(f^{(0)}f^{(0)})\gamma^{(0)}_{\mu\nu}\bigg)-\frac{L^2}{2}\hat\nabla^{(0)}_{(\mu} a_{\nu)}^{(2)}\,.
\end{align}
\par
Furthermore, expanding \eqref{emn} to the $O(z^4)$-order one obtains
\begin{align}
\gamma^{(6)}_{\mu\nu}=&-\frac{L^2}{3(d-6)}\bigg[\hat\nabla_\lambda\hat\gamma^\lambda_{(4)\mu\nu}-\frac{1}{2}\hat\nabla_{(\mu}\hat\nabla_{\nu)} \tr(m_{(4)})-\hat\nabla_\lambda(\hat\gamma^\sigma_{(2)\mu\nu}{m}_{(2)}^\lambda{}_\sigma)+\hat\nabla_{(\nu}(\hat\gamma^\sigma_{(2)\mu)\lambda}{m}_{(2)}^\lambda{}_\sigma)\\
&+\frac{1}{2}\hat\nabla_\sigma X^{(1)}\hat\gamma^\sigma_{(2)\mu\nu}-\hat\gamma^\sigma_{(2)\mu\lambda}\hat\gamma^\lambda_{(2)\sigma\nu}-\frac{2}{L^2}(m_{(2)}^3)^\rho{}_\nu\gamma^{(0)}_{\mu\rho}+\frac{8}{L^2}\gamma^{(4)}_{\rho(\mu}m_{(2)}^\rho{}_{\nu)}-\frac{1}{L^2}\gamma^{(4)}_{\mu\nu}X^{(1)}\nn\\
&-\frac{L^2}{2}f^{(0)}_{\sigma\mu}f^{(0)}_{\nu\rho}\gamma_{(2)}^{\rho\sigma}+L^2f^{(2)}_{\sigma(\mu}f^{(0)}_{\nu)\rho}\gamma_{(0)}^{\rho\sigma}-\frac{1}{L^2}\gamma^{(0)}_{\mu\nu}\Big(\tr(m_{(4)}m_{(2)})-\frac{1}{2}\tr(m_{(2)}^3)-\frac{L^4}{8}\tr(m_{(2)} f_{(0)}^2)\nn\\
&-\frac{L^4}{4}\hat\nabla_\rho a^{(2)}_\sigma f_{(0)}^{\rho\sigma}-\frac{L^2}{4}\hat\nabla_\sigma X^{(1)} a_{(2)}^\sigma+\frac{L^2}{2}\hat\nabla_\lambda(\gamma_{(2)}^{\lambda\rho} a^{(2)}_\rho)\Big)+2\hat\nabla_\lambda({m}_{(2)}^\lambda{}_{(\nu} a^{(2)}_{\mu)})-2\gamma^{(0)}_{\sigma(\nu}\hat\gamma^\sigma_{(2)\mu)\lambda}a_{(2)}^\lambda\nn\\
&-a^{(2)}_{(\nu}\hat\nabla_{\mu)} X^{(1)}-\hat\nabla_{(\nu}(X^{(1)} a^{(2)}_{\mu)})\bigg]-\frac{L^2}{3}\hat\nabla_{(\mu}a^{(4)}_{\nu)}-L^2a^{(2)}_{\mu}a^{(2)}_{\nu}+\frac{L^2}{6}a^{(2)}\cdot a^{(2)}\gamma^{(0)}_{\mu\nu}+\frac{L^2}{3}\hat\gamma^\lambda_{(2)\mu\nu}a^{(2)}_{\lambda}\nn\,,
\end{align}
where $f^{(2)}_{\sigma\mu}\equiv\hat\nabla_\sigma a^{(2)}_\mu-\hat\nabla_\mu a^{(2)}_\sigma$, and 
\begin{align}
\hat\gamma^\lambda_{(2)\mu\nu}&=\frac{1}{2}\gamma_{(0)}^{\lambda\rho}(\hat\nabla^{(0)}_\mu\gamma^{(2)}_{\nu\rho}+\hat\nabla^{(0)}_\nu\gamma^{(2)}_{\mu\rho}-\hat\nabla^{(0)}_\rho\gamma^{(2)}_{\mu\nu})=-\frac{L^2}{2}(\hat\nabla^{(0)}_\mu \hat P^\lambda{}_{\nu}+\hat\nabla^{(0)}_\nu\hat P_\mu{}^\lambda-\hat\nabla_{(0)}^\lambda\hat P_{\mu\nu})\,.
\end{align} 
(In the second step we used $\hat\nabla^{(0)}_\mu f^{(0)}_{\nu\rho}+\hat\nabla^{(0)}_\nu f^{(0)}_{\rho\mu}+\hat\nabla^{(0)}_\rho f^{(0)}_{\mu\nu}=0$.) The $\gamma^{(4)}_{\mu\nu}$ and $\gamma^{(6)}_{\mu\nu}$ above can be organized in to \eqref{g4} and \eqref{g6}, respectively.
\par
Finally, the $z\mu$-component of the Einstein equations gives
\begin{align}
0=&-\frac{L}{d-2}\frac{z^2}{L^2}\gamma_{(0)}^{\alpha\beta}\hat\nabla_\alpha^{(0)}\hat G^{(0)}_{\beta\mu}+L^{-1}\frac{z^4}{L^4}\bigg[\hat\nabla_\alpha\big (2m_{(4)\mu}^\alpha-(m_{(2)}^2)^\alpha{}_\mu\big)+\frac{1}{2}m_{(2)\mu}^\alpha\hat\nabla_\alpha X^{(1)}\nn\\
&+\frac{L^2}{2}\bigg(\hat\nabla\cdot\hat\nabla a_\mu^{(2)}-\hat\nabla_\mu\hat\nabla\cdot a^{(2)}+(R^{(0)}_{\beta\mu}+4f^{(0)}_{\beta\mu})\gamma_{(0)}^{\alpha\beta}a^{(2)}_\alpha-\hat\nabla_\alpha\big(f^{(0)}_{\beta\mu}m_{(2)\rho}^\alpha\gamma_{(0)}^{\rho\beta}\big)\nn\\
&-f^{(0)}_{\nu\rho}\gamma_{(0)}^{\alpha\nu}\hat\nabla_\alpha m_{(2)}^\rho{}_\mu+\frac{1}{2}f^{(0)}_{\beta\mu}\gamma_{(0)}^{\alpha\beta}\hat\nabla_\alpha X^{(1)}\bigg)-2\hat\nabla_\mu X^{(2)}+\frac{1}{2}\hat\nabla_\mu (X^{(1)})^2-\frac{1}{4}\hat\nabla_\mu\tr(m_{(2)}^2)
\bigg]+\cdots\nn\\
\label{emz}
&+\frac{z^d}{L^d}\bigg[\frac{d}{2L}\hat\nabla_\alpha n_{(0)\mu}^\alpha+\frac{L}{2}(\hat\nabla\cdot\hat\nabla p_\mu^{(0)}+\hat\nabla_\alpha\hat\nabla_\mu p_{(0)}^\alpha)\bigg]+\cdots\,.
\end{align}
One can observe that the $O(z^2)$-order of the above equation is exactly the contraction of the Weyl-Bianchi identity as shown in \eqref{BI}. By plugging in the results we got from the $zz$-equation, the $O(z^4)$-order can be organized into the identity \eqref{divWB}, which demonstrates the divergence of the Bach tensor. Also, the $O(z^d)$-order gives the conservation law of the improved energy-momentum tensor defined in \eqref{improvedT}.

\section{Expansions of the Raychaudhuri Equation and $\sqrt{-\det h}$}
\label{AppB}
Using the components of the Einstein equations \eqref{eomzz}--\eqref{eommn}, one can construct the following equation \cite{Ciambelli:2019bzz}:
\begin{align}
\label{eqa}
0&=\frac{g^{MN}(G_{MN}+\Lambda g_{MN})}{d-1}+(G_{zz}+\Lambda g_{zz})\nn\\
&=D_z\theta+L\nabla_\nu\varphi^\nu+L^2\varphi^2+\tr(\rho\rho)+\frac{L^2}{4}\tr(ff)-\frac{d}{L^2}\,,
\end{align}
where the indices $M,N$ represent the bulk components as $M=(z,\mu)$. This equation can be recognized as the Raychaudhuri equation of the congruence generated by $\un D_z$. Expanding each term in the above equation, we can write down a general expansion of this equation to any order. This combination of the components of the Einstein equations contains all the information we need for deriving $X^{(k)}$. We here provide some details of deriving $X^{(3)}$ and $X^{(4)}$ by means of the Raychaudhuri equation. 
\par
First, it is useful to expand the inverse of $h_{\mu\nu}$:
\begin{align}
\label{hinv}
h^{\mu\nu}(z;x)&=\frac{z^2}{L^2}\left[\gamma_{(0)}^{\mu\nu}(x)+\frac{z^2}{L^2}\gamma_{(2)}^{\mu\nu}(x)+...\right]+\frac{z^{d+2}}{L^{d+2}}\left[\pi_{(0)}^{\mu\nu}(x)+\frac{z^2}{L^2}\pi_{(2)}^{\mu\nu}(x)+...\right]\\
&=\frac{z^2}{L^2}\left[\gamma_{(0)}^{\mu\nu}(x)-\frac{z^2}{L^2}\tilde{m}_{(2)\rho}^{\mu}\gamma_{(0)}^{\rho\nu}(x)-\frac{z^4}{L^4}\tilde{m}_{(4)\rho}^{\mu}\gamma_{(0)}^{\rho\nu}(x)+\cdots\right]\nn+\frac{z^{d+2}}{L^{d+2}}\left[\tilde{n}_{(2)\rho}^{\mu}\gamma_{(0)}^{\rho\nu}(x)+\cdots\right]\,,
\end{align}
where $\tilde{m}_{(2k)\nu}^{\mu}\equiv-\gamma_{(2k)}^{\mu\rho}\gamma^{(0)}_{\rho\nu}$, $\tilde{n}_{(2k)\nu}^{\mu}\equiv-\pi_{(2k)}^{\mu\rho}\gamma^{(0)}_{\rho\nu}$. The above expansion can be solved order by order in terms of $m_{(2k)\nu}^{\mu}$ and $n_{(2k)\nu}^{\mu}$:
\begin{align}
\gamma_{(0)}^{\mu\nu}&=(\gamma^{(0)}_{\mu\nu})^{-1}\,,\qquad \tilde m_{(2)\nu}^{\mu}=m_{(2)\nu}^{\mu}\,,\qquad\tilde m_{(4)\nu}^{\mu}=m_{(4)\nu}^{\mu}-m_{(2)\rho}^{\mu}m_{(2)\nu}^{\rho}\,,\qquad\cdots\\
\tilde n_{(0)\nu}^{\mu}&=n_{(0)\nu}^{\mu}\,,\qquad\tilde n_{(2)\nu}^{\mu}=n_{(2)\nu}^{\mu}-m_{(2)\rho}^{\mu}n_{(0)\nu}^{\rho}-n_{(0)\rho}^{\mu}m_{(2)\nu}^{\rho}\,,\qquad\cdots.\nn
\end{align}
Also by taking the inverse of the metric, one finds the following relation:
\begin{align}
m_{(2p)}-\tilde{m}_{(2p)}=\sum^{p-1}_{k=1}\tilde{m}_{(2k)}m_{(2p-2k)}\,.
\end{align}
Specifically, we have
\begin{align}
m_{(2)}-\tilde{m}_{(2)}=0\,,\qquad m_{(4)}-\tilde{m}_{(4)}=m^2_{(2)}\,,\qquad m_{(6)}-\tilde{m}_{(6)}=m_{(2)}m_{(4)}+\tilde m_{(4)}m_{(2)}\,.
\end{align}
Now we expand the quantities defined in \eqref{quantities} to an arbitrary order by plugging the expansions \eqref{hex}, \eqref{aex} and \eqref{hinv} into their definitions. For the purpose of finding the Weyl anomaly, here we only keep the $m_{(2p)}$ and $a_{(2p)}$ terms in the first series of $h_{\mu\nu}$ and $a_\mu$ and neglect the $n_{(2p)}$ and $p_{(2p)}$ terms. The expansions of these quantities are

\begin{align}
\label{exp1}
\rho^\mu{}_\nu=&-\delta^\mu{}_\nu+\frac{1}{2}\sum_{p=1}^\infty\left(\frac{z}{L}\right)^{2p}\Big[p(m_{(2p)}+\tilde{m}_{(2p)})+\sum_{k=1}^{p-1}(2k-p)\tilde{m}_{(2k)}m_{(2p-2k)}\Big]^\mu{}_\nu+O(z^d)\,,\\
\theta=&-\frac{d}{L}+\frac{1}{2L}\sum^{\infty}_{p=1}\bigg(\frac{z}{L}\bigg)^{2p}\bigg[p\tr(m_{(2p)}+\tilde{m}_{(2p)})+\sum^{p-1}_{k=1}(2k-p)\tr\tilde{m}_{(2k)}m_{(2p -2k)}\bigg]+O(z^d)\,,
\end{align}
\begin{align}
\varphi_\mu={}&\frac{1}{L}\sum^\infty_{p=0}\left(\frac{z}{L}\right)^{2p}2pa_\mu^{(2p)}+O(z^{d-2})\,,\\
f_{\mu\nu}=&\sum_{p=0}^{\infty}\left(\frac{z}{L}\right)^{2p}\big[f^{(2p)}_{\mu\nu}+\sum_{q=1}^{p-1} 2q(a_\mu^{(2p-2q)} p_\nu^{(2q)}-a_\nu^{(2p-2q)} p_\mu^{(2q)})\big]+O(z^{d-2})\,,\\
\gamma^\lambda{}_{\mu\nu}={}&\gamma^\lambda_{(0)\mu\nu}-\sum_{p=1}^\infty\left(\frac{z}{L}\right)^{2p}\bigg(\sum_{q=0}^{p-1}\tilde{m}_{(2q)}^\lambda{}_\rho\hat\gamma^\rho_{(2p-2q)\mu\nu}+\frac{1}{2}\sum_{q=0}^{p-1}[\tilde{m}_{(2q)}\gamma_{(0)}^{-1}]^{\lambda\rho}\sum_{k=0}^{p-q-1}(2k-2)\qquad\nn\\
\label{exp5}
&\times(a^{(2p-2q-2k)}_\mu\gamma^{(2k)}_{\nu\rho}+a^{(2p-2q-2k)}_\nu\gamma^{(2k)}_{\mu\rho}-a^{(2p-2q-2k)}_\rho\gamma^{(2k)}_{\mu\nu})\bigg)+O(z^{d-2})\,,
\end{align}
where 
\begin{align*}
f^{(0)}_{\mu\nu}&=\p_\mu a_\nu^{(0)}-\p_\nu a_\mu^{(0)}\,,\qquad f^{(2k)}_{\mu\nu}=\hat\nabla^{(0)}_\mu a_\nu^{(2k)}-\hat\nabla^{(0)}_\nu a_\mu^{(2k)}\quad(k>0)\,,\\
\gamma^\lambda_{(0)\mu\nu}&=\frac{1}{2}\gamma_{(0)}^{\lambda\rho}\big(
 \p_\mu \gamma^{(0)}_{\nu\rho}
 +\p_\nu \gamma^{(0)}_{\mu\rho}-\p_\rho \gamma^{(0)}_{\mu\nu}\big)-\big(a^{(0)}_\mu\delta^\lambda{}_\nu+a^{(0)}_\nu\delta^\lambda{}_\mu-a^{(0)}_\rho\gamma_{(0)}^{\lambda\rho}\gamma^{(0)}_{\mu\nu}\big)\,,\\
\hat\gamma^\lambda_{(2k)\mu\nu}&=\frac{1}{2}\gamma_{(0)}^{\lambda\rho}(\hat\nabla^{(0)}_\mu\gamma^{(2k)}_{\nu\rho}+\hat\nabla^{(0)}_\nu\gamma^{(2k)}_{\mu\rho}-\hat\nabla^{(0)}_\rho\gamma^{(2k)}_{\mu\nu})\quad(k>0)\,.
\end{align*} 
Expanding everything in \eqref{eqa} using \eqref{exp1}--\eqref{exp5}, we obtain the following equation:
\begin{align}
0={}&\frac{1}{L^2}p(p-1)\tr(m_{(2p)}+\tilde{m}_{(2p)})+\frac{1}{L^2}\sum^{p-1}_{q=1}(p-1)(2q-p)\tr\tilde{m}_{(2q)}m_{(2p-2q)}\nn\\
&-\sum^{p-1}_{q=1}2q\hat\nabla_\mu a_\nu^{(2q)}\big[\tilde{m}_{(2p-2q-2)}\gamma_{(0)}^{-1}\big]^{\mu\nu}-\sum^{p-1}_{q=1}\sum^{q-1}_{k=0}(2p-2q+2k)2ka_\mu^{(2p-2q)}a_\nu^{(2k)}\big[\tilde{m}_{(2q-2k-2))}\gamma_{(0)}^{-1}\big]^{\mu\nu}\nn\\
&-\sum^{p-1}_{q=1}\sum_{k=0}^{q-1}\sum_{n=0}^{p-q-1}na_\lambda^{(2n)}[\tilde{m}_{(2p-2q-2n-2)}\gamma^{-1}_{(0)}]^{\mu\nu}\bigg(\tilde{m}_{(2k)}^\lambda{}_\rho\hat\gamma^\rho_{(2q-2k)\mu\nu}\nn\\
&\quad-[\tilde{m}_{(2k)}\gamma_{(0)}^{-1}]^{\lambda\rho}\sum_{m=0}^{q-k-1}(2-2m)(a^{(2q-2k-2m)}_\mu\gamma^{(2m)}_{\nu\rho}+a^{(2q-2k-2m)}_\nu\gamma^{(2m)}_{\mu\rho}-a^{(2q-2k-2m)}_\rho\gamma^{(2m)}_{\mu\nu})\bigg)\nn\\
&+\frac{1}{4L^2}\sum_{q=1}^{p-1}(p-q)\tr\bigg[(m_{(2p-2q)}+\tilde{m}_{(2p-2q)})\Big[q(m_{(2q)}+\tilde{m}_{(2q)})+\sum_{k=1}^{q-1}2(2k-q)\tilde{m}_{(2k)}m_{(2q-2k)}\Big]\bigg]\nn\\
&+\frac{1}{4L^2}\sum_{q=1}^{p-1}\sum_{k=1}^{q-1}\sum_{m=1}^{p-q-1}(2k-q)(2m-p+q)\tr\big[\tilde{m}_{(2k)}m_{(2q-2k)}\tilde{m}_{(2m)}m_{(2p-2q-2m)}\big]\nn\\
&+\frac{L^2}{4}\sum_{q=1}^{p-1}\sum_{k=0}^{q-1}\big[f^{(2k)}_{\mu\rho}+\sum_{m=1}^{k-1} 2m(a_\mu^{(2k-2m)} a_\rho^{(2m)}-a_\rho^{(2k-2m)} a_\mu^{(2m)})\big][\tilde{m}_{(2q-2k-2)}\gamma^{-1}_{(0)}]^{\rho\nu}\nn\\
\label{Rayex}
&\quad\times\sum_{n=0}^{p-q-1}\big[f^{(2n)}_{\nu\sigma}+\sum_{s=1}^{n-1} 2s(a_\nu^{(2n-2s)} a_\sigma^{(2s)}-a_\sigma^{(2n-2s)} a_\nu^{(2s)})\big][\tilde{m}_{(2p-2q-2n-2)}\gamma^{-1}_{(0)}]^{\sigma\mu}\,.
\end{align}
From this equation, one can find $\tr(m_{(2p)}+\tilde{m}_{(2p)})$ in terms of $m_{(2q)}$ and $\tilde{m}_{(2q)}$ for all $q<p$. 
\par
Taking $p=3$ we get the Raychaudhuri equation at the $O(z^6)$-order:
\begin{align}
0={}&\frac{6}{L^2}\tr(m_{(6)}+\tilde{m}_{(6)})+\frac{4}{L^2}\tr(m_{(4)}m_{(2)})-\frac{4}{L^2}\tr(m_{(2)}^3)-\frac{L^2}{2}m_{(2)}^\mu{}_\alpha f^\alpha_{(0)\beta} f^\beta_{(0)\mu}\nn\\
\label{Ray6d}
&+4\hat\nabla\cdot a^{(4)}-2m_{(2)}^\mu{}_\rho\gamma_{(0)}^{\rho\nu}\hat\nabla_\nu a_\mu^{(2)}-2\gamma_{(0)}^{\mu\nu}\hat\gamma^{\lambda}_{(2)\mu\nu}a^{(2)}_\lambda-2(d-6)a^2_{(2)}+\frac{L^2}{2}f^{(2)}_{\mu\nu}f_{(0)}^{\nu\mu}\,.
\end{align}
And for $p=4$, we have the Raychaudhuri equation at the $O(z^8)$-order: 
\begin{align}
0={}&\frac{12}{L^2}\tr(m_{(8)}+\tilde{m}_{(8)})+\frac{6}{L^2}\tr(m_{(6)}m_{(2)})-\frac{22}{L^2}\tr(m_{(4)}m_{(2)}^2)+\frac{9}{L^2}\tr(m_{(2)}^4)+\frac{4}{L^2}\tr(m_{(4)}^2)\nn\\
&+\frac{L^2}{4}f^{(0)}_{\mu\rho}f_{(0)}^{\nu\sigma}m_{(2)}^\rho{}_\nu m_{(2)}^\mu{}_\sigma+\frac{L^2}{2}f^{(0)}_{\mu\rho}f_{(0)}^{\rho\sigma}(m_{(2)}^2)^\mu{}_\sigma-\frac{L^2}{2}f^{(0)}_{\mu\rho}f_{(0)}^{\rho\sigma}(m_{(4)})^\mu{}_\sigma+6\hat\nabla\cdot a^{(6)}\nn\\
&-4\hat\nabla_\mu a_\nu^{(4)}\gamma_{(2)}^{\mu\nu}+L^2\hat\nabla_{[\mu}a^{(4)}_{\rho]}f_{(0)}^{\rho\mu}-4a_\sigma^{(4)}\gamma_{(0)}^{\mu\nu}\hat\gamma_{(2)}^\sigma{}_{\mu\nu}-6(d-8)a^{(4)}\cdot a^{(2)}-2\hat\nabla_\mu a_\nu^{(2)}\gamma_{(4)}^{\mu\nu}\nn\\
&-2a_\sigma^{(2)}\gamma_{(0)}^{\mu\nu}\hat\gamma_{(4)}^\sigma{}_{\mu\nu}+2\hat\nabla_\mu a_\nu^{(2)}(m_{(2)}^2)^\mu{}_\rho\gamma_{(0)}^{\rho\nu}+L^2\hat\nabla_{[\mu}a^{(2)}_{\rho]}\hat\nabla^{[\rho}a_{(2)}^{\mu]}-2L^2\hat\nabla_{[\mu}a^{(2)}_{\rho]}f_{(0)}^{\rho\sigma}m_{(2)}^\mu{}_\sigma\nn\\
\label{Ray8d}
&+2a_\sigma^{(2)}\gamma_{(2)}^{\mu\nu}\hat\gamma_{(2)}^\sigma{}_{\mu\nu}+2a_\lambda^{(2)}\gamma_{(0)}^{\mu\nu}m_{(2)}^\lambda{}_\sigma\hat\gamma_{(2)}^\sigma{}_{\mu\nu}+2(d-8)a^{(2)}_\mu a^{(2)}_\nu\gamma_{(2)}^{\mu\nu}+2X^{(1)}a^{(2)}\cdot a^{(2)}\,.
\end{align}
\par
Now let us look at the expansion of $\sqrt{-\det h}$. Using the fact that $\theta=D_z(\ln \sqrt{-\det h})$, we can write down the expansion of $\sqrt{-\det h}$ to any order as
\begin{align}
\label{deth}
\sqrt{-\det h}={}&\sqrt{-\det\gamma_{(0)}}\bigg(\frac{z}{L}\bigg)^{-d}\sum_0^\infty\frac{1}{n!}\\
&\times\left[\frac{1}{2}\sum^{\infty}_{m=1}\bigg(\frac{z}{L}\bigg)^{2m}\bigg[\frac{1}{2}\tr(m_{(2m)}+\tilde{m}_{(2m)})+\sum^{m-1}_{k=1}\bigg(\frac{k}{m}-\frac{1}{2}\bigg)\tr(\tilde{m}_{(2k)}m_{(2m-2k)})\bigg]\right]^n\nn\,.
\end{align}
Comparing with \eqref{sqrth}, at the $O(z^6)$-order and the $O(z^8)$-order, the above equation gives respectively
\begin{align}
\label{X3}
X^{(3)}={}&\frac{1}{2}\tr(m_{(6)}+\tilde{m}_{(6)})-\frac{1}{6}\tr (m_{(2)}^3)+\frac{1}{2}X^{(1)}X^{(2)}-\frac{1}{12}(X^{(1)})^3\,,\\
X^{(4)}={}&\frac{1}{2}\tr(m_{(8)}+\tilde{m}_{(8)})-\frac{1}{2}\tr(m_{(4)}m_{(2)}^2)+\frac{1}{4}\tr(m_{(2)}^4)\nn\\
\label{X4}
&+\frac{1}{2}X^{(3)}X^{(1)}-\frac{1}{4}X^{(2)}(X^{(1)})^2+\frac{1}{4}(X^{(2)})^2+\frac{1}{32}(X^{(1)})^{4}\,.
\end{align}
Now solving for $\tr(m_{(6)}+\tilde{m}_{(6)})$ from \eqref{Ray6d} and plugging \eqref{g2}, \eqref{g4} and \eqref{X1X2} into \eqref{X3}, we can organize all the $m_{(2)}$ and $f_{(0)}$ terms in $X^{(3)}$ and get \eqref{X3P}. Similarly, plugging $\tr(m_{(8)}+\tilde{m}_{(8)})$ obtained from \eqref{Ray8d} into \eqref{X4}, the expression for $X^{(4)}$ can be organized in terms of the Weyl-Schouten tensor and extended Weyl-obstruction tensors as
\begin{align}
\frac{24}{L^2}X^{(4)}={}&L^6\bigg(\frac{1}{8}\hat P^4-\frac{3}{4}\tr(\hat P^2)\hat P^2+\frac{3}{8}[\tr(\hat P^2)]^2+\tr(\hat P^3)\hat P-\frac{3}{4}\tr(\hat P^4)-\tr(\hat\Omega_{(1)}\hat P)\hat P+\tr(\hat\Omega_{(1)}\hat P^2)\nn\\
&-\frac{1}{4}\tr(\hat\Omega_{(1)}^2)-\frac{1}{4}\tr(\hat\Omega_{(2)}\hat P)\bigg)+2(d-8)\big[3a^{(4)}\cdot a^{(2)}+a^{(2)}_\mu a^{(2)}_\nu(\hat P^{\mu\nu}-\hat P\gamma_{(0)}^{\mu\nu})\big]-6\hat\nabla\cdot a^{(6)}\nn\\
&-L^2\hat\nabla_\mu \big[a_\nu^{(4)}(4\hat P^{\mu\nu}+2\hat P^{\nu\mu}-4\hat P\gamma_{(0)}^{\mu\nu})\big]-\frac{L^2}{2}\hat\nabla_\mu \big[a^{(2)}_{\nu}(3\hat\nabla^{\nu}a_{(2)}^{\mu}+\hat\nabla^{\mu} a^{\nu}_{(2)}-3\hat\nabla\cdot a_{(2)}\gamma_{(0)}^{\mu\nu})\big]\nn\\
&+L^4\hat\nabla_\mu\big[a^{(2)}_\nu(3\hat P^{\mu\nu}\hat P+\hat P^{\nu\mu}\hat P)\big]+\frac{3L^4}{2}\hat\nabla^\mu\big[a^{(2)}_\mu(\tr(\hat P^2)-\hat P^2) \big]-\frac{3L^4}{2}\hat\nabla_\mu (a_\nu^{(2)}\hat\Omega_{(1)}^{\mu\nu})\nn\\
\label{X8t}
&-\frac{L^4}{4}\hat\nabla_{\mu}\big[a_{\nu}^{(2)}(3\hat P^{\rho\mu}\hat P^\nu{}_\rho-5\hat P^{\rho\mu}\hat P_\rho{}^\nu+7\hat P^{\mu\rho}\hat P_\rho{}^\nu-9\hat P^{\mu\rho}\hat P^\nu{}_\rho)\big]\,,
\end{align}
which leads to \eqref{X4P}.

\section{Dilatation Operator Method}
\label{AppC}
In this appendix we will derive the holographic Weyl anomaly in $8d$ using the recursive algorithm of  \cite{Anastasiou:2020zwc} which we will refer to as the {\it dilatation operator method}. We point out that this method uses the usual FG gauge where the Weyl connection is turned off. This alternative way of finding the Weyl anomaly will provide a nontrivial consistency check of the results presented in Section \ref{Sec5}. For completeness we start with a brief review of the algorithm of the dilatation operator method. We then apply the algorithm one step further than \cite{Anastasiou:2020zwc} and  compute the holographic Weyl anomaly in $8d$. \subsection{Review of the Algorithm}\label{subsection_C1}
We start by using the metric in the FG gauge  \eqref{FG}
\begin{equation}
\td s^2=\td r^2+h_{\mu\nu}(r;x)\td x^\mu \td x^\nu\,,\qquad\mu,\nu=1,\cdots,d\,,
\end{equation}
where we changed the coordinates by setting $r=- L\text{ln}\left(z/L\right)$. The Einstein-Hilbert action in the bulk manifold $M$ with a Gibbons-Hawking boundary term is
\begin{equation}
S_{\text{EH-GH}}= \frac{1}{2\kappa^2} \int_{M}\td r\td ^{d}x\sqrt{-\text{det}g}(R-2\Lambda)+ \frac{1}{\kappa^2} \int_{\partial M_{r_{c}}}\td ^{d}x\sqrt{-\det h}K\,,
\end{equation}
where $\kappa^2 =8\pi G$, $\Lambda= -\frac{d(d-1)}{2L^2}$, and $\partial M_{r_{c}}$ is a cutoff surface at some large value of $r_{c}$.\footnote{We abuse notation and call the cuttof surface $\partial M_{r_{c}}$ even though it is not necessarily the boundary of $M$. } Taking a metric variation and evaluating the result on-shell one gets
\begin{equation}
\delta S_{EHGH}^{\text{ o.s}}= \int_{\partial M_{r_{c}}} \td ^{d}x\pi^{\mu\nu}\delta h_{\mu\nu}\,,\qquad\pi^{\mu\nu}= \frac{1}{2\kappa^2}\sqrt{-\det h}\left(K h^{\mu\nu}- K^{\mu\nu}\right)\,,
\end{equation}
where $K_{\mu\nu}= \frac{1}{2}\partial_{r}h_{\mu\nu}$ is the extrinsic curvature tensor in the FG gauge, and $h_{\mu\nu}$ is the induced metric on $\partial M_{r_{c}}$. The boundary tensor density $\pi^{\mu\nu}$ appears in many different contexts; it was first defined in \cite{brown1993quasilocal} and was later used in \cite{Balasubramanian_1999} to define a boundary stress tensor in an asymptotically AdS spacetime (with the inclusion of necessary counterterms to cancel divergences). It also appears as the  conjugate momenta in the ADM formalism \cite{arnowitt1960canonical}.\footnote{The sign difference in the definition of $\pi^{\mu\nu}$ in \cite{arnowitt1960canonical} arises because here we consider the radial evolution which is in a spacelike direction.}

The $rr$- and $r\mu$-components of the Einstein equations, i.e.\ $G^{rr}+ \Lambda g^{rr}=0$ and $G^{r\mu}=0$, can be written in terms of the conjugate momenta as follows:
\begin{align}\label{EE_p1}
\frac{2\kappa^2}{\sqrt{-\det h}}\left(\pi\indices{^{\mu}_{\nu}}\pi^{\nu}_{\mu}- \frac{1}{d-1}\pi^2\right)+ \frac{1}{2\kappa^2}\sqrt{-\det h}\left(\mathring R -2\Lambda\right)&=0\,,\\
\label{EE_p2}
\mathring\nabla_{\rho}\pi^{\rho\mu}&=0\,,
\end{align}
where $\pi\equiv h_{\mu\nu}\pi^{\mu\nu}$ ,$\mathring \nabla$ is the LC connection of the induced metric $h_{\mu\nu}$ on $\partial M_{r_{c}}$, and $\mathring R$ is the LC Ricci scalar of $h_{\mu\nu}$. Note that the indices are raised and lowered using the induced metric $h_{\mu\nu}$. Eq.\ $\eqref{EE_p1}$ and \eqref{EE_p2} are the well-known Hamiltonian and momentum constraints in the ADM language \cite{arnowitt1960canonical}. 
\par
The dilatation operator method of solving the constraint equations uses an asymptotic expansion of the conjugate momenta in terms of the induced metric.
One assumes a Hamilton-Jacobi functional $S[h]$ such that
\begin{equation}\label{pi_def}
\pi^{\mu\nu}= \frac{\delta S[h]}{\delta h_{\mu\nu}}\,,\qquad S[h]\equiv \int_{\partial M_{r_{c}}}  \td ^{d}x \mathcal{L}[h]\,,
\end{equation}
where $S[h]$ is a local diffeomorphic invariant functional of the induced metric.\footnote{In \cite{Anastasiou:2020zwc} the functional $S[h]$ is used to derive the boundary terms for a well-defined bulk variational problem. Since these are tangential remarks for the calculation of the Weyl anomaly, we simply neglect them and refer the reader there for more details.}
\par
The momentum constraint \eqref{EE_p2} is now trivially satisfied. The Hamiltonian constraint \eqref{EE_p1} can be solved asymptotically by writing
\begin{equation}\label{L_expansion}
\mathcal{L}= \sum_{k=0}\mathcal{L}_{(2k)}[h]\,,\qquad\delta_{D}\mathcal{L}_{(2k)}[h]= (d-2k)\mathcal{L}_{(2k)}[h]\,,
\end{equation}
where $\delta_{D}$ is the dilatation operator \cite{Papadimitriou:2004ap} (acting on metric functionals), defined as
\begin{align}
\label{deltaD}
\delta_{D}\equiv \int \td ^{d}x \left[2 h_{\mu\nu}\frac{\delta }{\delta h_{\mu\nu}}\right]\,.
\end{align}
It is useful to keep in mind that for an AlAdS spacetime the radial derivative $\partial_{r}$ asymptotes the dilatation operator, i.e.\ $\delta_D \sim  \frac{\partial}{\partial r}$ (see \cite{Anastasiou:2020zwc}). We can view the dilatation operator expansion as another asymptotic expansion near the conformal boundary since $h_{\mu\nu}\sim e^{2r/L}\gamma_{\mu\nu}^{(0)}+ \cdots$.
\par
The expansion \eqref{L_expansion} together with \eqref{pi_def} implies an expansion of $\pi^{\mu\nu}$ in terms of the dilatation weight:
\begin{equation}\label{pi_2n- definition}
 \pi^{\mu\nu}_{(2k)}= \frac{\delta}{\delta h_{\mu\nu}}\int_{\partial M_{r_{c}}}\td^{d}x\mathcal{L}_{(2k)}\,,\qquad\pi^{\mu\nu}= \sum_{k=0}\pi^{\mu\nu}_{(2k)}\,.
\end{equation}
The $\mathcal{L}_{(2k)}[h]$ are defined only up to total derivative terms in $\partial M_{r_{c}}$. 
To set up the recursive algorithm we need the following relation:
\begin{equation}\label{trace_of_pi}
\pi_{(2k)}=\frac{L}{2} \mathcal{Q}_{(2k)}\,,\qquad\pi_{(2k)}:= h_{\mu\nu}\pi^{\mu\nu}_{(2k)}\,,
\end{equation}
where we defined the \emph{Q-curvature} 
\begin{align}
\mathcal{Q}_{(2k)} := \frac{d-2k}{L}\mathcal{L}_{(2k)}\,.
\end{align} 
The Q-curvature is the main quantity we are interested in, since it is proportional to the Weyl anomaly in a specific even dimension [see \eqref{QandA}]. Plugging $\eqref{pi_2n- definition}$ into $\eqref{EE_p1}$ and making use of $\eqref{trace_of_pi}$ we find
\begin{equation}\label{Q-Algorithm}
\mathcal{Q}_{(2k)}= \frac{2\kappa^2}{\sqrt{-\det h}}\sum_{m=1}^{k-1}\left(\pi\indices{_{(2m)}^{\mu}_{\nu}}\pi\indices{_{(2k-2m)}^{\nu}_{\mu}}- \frac{1}{d-1}\pi_{(2m)}\pi_{(2k-2m)}\right)\,,
\end{equation}
 where we used the initial values
\begin{equation}\label{Initial values}
\mathcal{Q}_{(2)}= \frac{\sqrt{-\det h}}{2\kappa^2}\mathring R\,,\qquad\pi_{(0)}^{\mu\nu}=\frac{(d-1)}{2\kappa^2 L}\sqrt{-\det h}h^{\mu\nu}\,.
\end{equation}
\par
Equations $\eqref{Q-Algorithm}$, $\eqref{pi_2n- definition}$ and the initial values \eqref{Initial values}, are enough to fix the iterative algorithm of the dilatation operator method. Expanding on this a little more, given the value of $\mathcal{Q}_{(2)}$ we can use \eqref{pi_2n- definition} to find $\pi_{(2)}^{\mu\nu}$.  Since we now have $\pi_{(2)}^{\mu\nu}$ and $\pi_{(0)}^{\mu\nu}$, we can find $\mathcal{Q}_{(4)}$ from \eqref{Q-Algorithm}. The process can then be iterated to compute $\pi_{(2k)}^{\mu\nu}$ and $\mathcal{Q}_{(2k)}$ to any order. The recursive algorithm has been solved up to $\pi_{(4)}^{\mu\nu}$ and $\mathcal{Q}_{(6)}$ in \cite{Anastasiou:2020zwc}. In the next section we will push the calculation one step further for finding $\pi^{\mu\nu}_{(6)}$ and $\mathcal{Q}_{(8)}$. This will enable us to find the Weyl anomaly in $8d$.

\subsection{Results and Anomaly in $8d$}
We have explained the algorithm for solving the Hamiltonian and momentum constraints. We now focus on the Q-curvature, which is expressed in terms of the conjugate momenta in \eqref{Q-Algorithm}. The  Weyl anomaly in $d$-dimension corresponds to the Q-curvature for $d=2k$\cite{Anastasiou:2020zwc}:
\begin{equation}\label{QandA}
{\cal A}_k =- L\int \td ^{d}x \ln {\cal B}{\cal Q}_{(2k)}^{d=2k}\,.
\end{equation}
\par
We now present the results of the algorithm presented in \ref{subsection_C1}. First, we review the results for $\pi_{(2k)}^{\mu\nu}$ and $\mathcal{Q}_{(2k)}$ up to $k=3$:
\begin{equation}\label{old_pi4}
\begin{split}
\pi_{(0)}^{\mu\nu}&=\frac{(d-1)}{2\kappa^2 L}\sqrt{-\det h}h^{\mu\nu}\,,\\
\pi_{(2)}^{\mu\nu}&= -\frac{\sqrt{-\det h}L}{2\kappa^2}\left(\mathring P^{\mu\nu}- \mathring Ph^{\mu\nu}\right)\,,\\
\pi_{(4)}^{\mu\nu}&=- \frac{\sqrt{-\det h}L^3}{2\kappa^2(d-4)(d-2)}\left[\mathring B^{ij}+ (d-4)\left(\mathring P^{\mu}_{~\lambda}\mathring P^{\lambda\nu}- \mathring P \mathring P^{\mu\nu} - \frac{1}{2}h^{\mu\nu}(\tr (\mathring P^{2}) -\mathring P^2)\right)\right]\,,
\end{split}
\end{equation}
and
\begin{equation}\label{old_Q6}
\begin{split}
\mathcal{Q}_{(2)}&= \frac{\sqrt{-\det h}}{2\kappa^2}\mathring{R}\,,\\
\mathcal{Q}_{(4)}&= \frac{\sqrt{-\det h}L^2}{2\kappa^2}\left[\tr (\mathring P^2)- \mathring P^2\right]\,,\\ 
\mathcal{Q}_{(6)}&=\frac{\sqrt{-\det h}L^4}{\kappa^2(d-4)(d-2)}\left[\tr (\mathring P \mathring B)+ (d-4)\bigg(\tr (\mathring P^3 )-  \frac{3}{2}\mathring P \tr (\mathring P^2)+ \frac{1}{2}\mathring P^3\bigg)  \right]\,.
\end{split}
\end{equation}
We can see that $\mathcal{Q}_{(2)}$, $\mathcal{Q}_{(4)}$ and $\mathcal{Q}_{(6)}$ correspond to the LC counterparts of the Weyl anomaly shown in \eqref{2dAP}, \eqref{4dAP} and \eqref{6dA}, respectively. Using $\mathcal{Q}_{(6)}$ in \eqref{old_Q6} we can calculate $\pi_{(6)}^{\mu\nu}$ from \eqref{pi_2n- definition} by taking a metric variation of $\mathcal{Q}_{(6)}$. The result is as follows:
\begin{align}\label{pi6final_2}
\pi_{(6)}^{\mu\nu}={}& -\frac{L^5}{\kappa^2 (d-6)(d-4)(d-2)^2}\bigg[ \mathcal{\mathring O}_{(6)}^{\mu\nu}
+(d-6) \bigg(\mathring P\indices{^{(\mu}_{\lambda}}\mathring B^{\nu)\lambda}- \mathring P \mathring B^{\mu\nu} -2\mathring P_{\rho\lambda}\mathring \nabla^\lambda \mathring C^{(\mu\nu)\rho}\nn\\& 
 - (\mathring\nabla_\lambda P)\mathring C^{(\mu\nu)\lambda} + \mathring C^{\rho \mu \lambda}\mathring C\indices{_{\lambda}^{\nu}_{\rho}}- \frac{1}{2}\mathring C\indices{^{\mu}^{\rho\lambda}}\mathring C\indices{^{\nu}_{\rho\lambda}}+ \mathring P \mathring W\indices{^{\mu}_{\alpha\beta}^{\nu}}\mathring P^{\alpha\beta}+ \frac{1}{2}\mathring\nabla^{\lambda}\mathring\nabla_{\lambda} (\mathring P^{\mu\beta}\mathring P\indices{^{\nu}_{\beta}} - \mathring P\mathring P^{\mu\nu}) \nn\\&
 - \frac{(d-2)}{4(d-1)} \mathring\nabla^{\mu}\mathring\nabla^{\nu}(\tr (\mathring P^2)- \mathring P^2)-\frac{1}{4(d-1)}h^{\mu\nu}\mathring\nabla^{\lambda}\mathring\nabla_{\lambda} (\tr (\mathring P^2)- \mathring P^2)+(d-4) \mathring P^{\mu}_{~~\alpha}\mathring P^{\alpha \beta}\mathring P\indices{_{\beta}^{\nu}}
\nn\\&-  \frac{(3d^2 -12d+8)}{4(d-1)} \mathring P^{\mu\nu}(\tr (\mathring P^2)-\mathring P^2)- (d-4)\mathring P \mathring P^{\mu}{}_{\lambda}\mathring P^{\lambda \nu}-\frac{1}{2}h^{\mu\nu}\Big(\tr (\mathring P\mathring B)+ (d-4)\text{tr}(\mathring P^3)\nn \\&
 - \frac{(3d^2-14d+10)}{2(d-1)}\mathring P\text{tr}(\mathring P^2) + \frac{(d^2 -4d+2)}{2(d-1)}\mathring P^3\Big)\bigg) \bigg]\,,
\end{align}
where ${\cal \mathring O}^{\mu\nu}_{(6)}$ is the LC obstruction tensor defined in  \eqref{O6}. We have also checked that $\pi_{(6)}^{\mu\nu}$ is divergence-free in any dimension, as is required by \eqref{EE_p2}.\footnote{This calculation was done thanks to the Mathematica package \href{https://people.brandeis.edu/~headrick/Mathematica/index.html}{diffgeo.m} by Matthew Headrick.}

By plugging \eqref{pi6final_2} and \eqref{old_pi4} into \eqref{Q-Algorithm} we find after some reorganization the expression of $\mathcal{Q}_{(8)}$ as follows:
\begin{align}\label{Q8_Final_4}
\mathcal{Q}_{(8)}={}&  \frac{2L^6 \sqrt{-\det h}}{\kappa^2(d-6)(d-4)(d-2)^2}\bigg[ \frac{(d-2)}{2(d-4)}\tr (\mathring P \mathring{\mathcal{O}}_{(6)})+(d-6)\bigg(\frac{3}{4(d-4)}\tr (\mathring B^2) + \frac{5d-16}{2 (d-4)}\tr (\mathring P^2 \mathring B) \nn\\&
-\frac{5d-16}{2(d-4)}\mathring P \tr (\mathring P\mathring B)-\frac{5d^2 -20d+8}{16(d-1)}\mathring P^4+ \frac{16-5d}{2}\mathring P\tr (\mathring P^3)+\frac{15d^2 -62d+40}{8(d-1)}\mathring P^2 \tr (\mathring P^2)\nn\\&
- \frac{13d^2 -44d+24}{16(d-1)}(\tr (\mathring P^2) )^2+\frac{7d-20}{4}\tr (\mathring P^4)  \bigg)
 +\mathring\nabla_{\mu}K^{\mu} \bigg]\,,
\end{align}
where $K^{\mu}=\frac{(d-6)}{2(d-4)} K_{0}^{\mu}+ \frac{(d-6)}{2}K_{1}^{\mu}+ \frac{(d-6)}{2}K_{2}^{\mu}+ \frac{(d-6)(d-2)}{4(d-1)}K_{3}^{\mu} $, with
\begin{equation}
\begin{split}
K_{0}^{\mu}&:= \mathring P^{\alpha\beta}\LCnabla^{\mu}\mathring B_{\alpha\beta}-\mathring B_{\alpha\beta}\mathring\nabla^{\mu}\mathring P^{\alpha\beta}+ \mathring B^{\mu \beta}\mathring\nabla_{\beta}\mathring P - \mathring P\mathring\nabla_{\alpha}\mathring B^{\alpha \mu}\,,\\
K_{1}^{\mu}&:=(\mathring P^{\mu \lambda}\mathring P^{\beta}_{\lambda}-\mathring P\mathring P^{\mu \beta})\mathring\nabla_{\beta}\mathring P - \mathring P\mathring\nabla_{\beta}(\mathring P^{\beta \alpha}\mathring P^{\mu}_{\alpha}- \mathring P\mathring P^{\beta \mu})\,,\\
K_{2}^{\mu}&:= \mathring P_{\alpha\beta}\mathring\nabla^{\mu}(\mathring P^{\alpha \lambda}\mathring P\indices{_{\lambda}^{\beta}} - \mathring P\mathring P^{\alpha\beta}) - (\mathring\nabla^{\mu}\mathring P_{\alpha\beta})(\mathring P^{\alpha \lambda}\mathring P_{\lambda}^{~\beta}- \mathring P\mathring P^{\alpha\beta})\,,\\
K_{3}^{\mu}&:=(\tr (\mathring P^2)- \mathring P^2)\mathring\nabla^{\mu}\mathring P - \mathring P^{\mu \lambda}\mathring\nabla_{\lambda}(\tr (\mathring P^2) - \mathring P^2)\,.
\end{split}
\end{equation}
If we plug $d=8$ into \eqref{Q8_Final_4} , we find that the holographic Weyl anomaly $\mathcal{A}_{4}$ in $8d$ is
\begin{align}
\label{8dAC}
\mathcal{A}_{4}= -L\int \td^{8}x \ln {\cal B}\mathcal{Q}_{(8)}^{d=8}=&  -\int \td^{8}x\frac{L^7 \sqrt{-\det h}}{48\kappa^2}\bigg[\frac{1}{4}\tr(\mathring P\mathring{\cal O}_{(6)}) + \frac{1}{8}\tr(\mathring B^2)+2\tr(\mathring P^2 \mathring B){}- 2\mathring P\tr(\mathring P\mathring B){}\nn\\
&+ 6 \tr(\mathring P^4)- 3\tr(\mathring P^2)^{2}+ 6\mathring P^2 \tr(\mathring P^2) - 8\mathring P\tr(\mathring P^3) - \mathring P^4 + \LCnabla_{\mu}K^\mu\bigg]\,.
\end{align}
This result agrees with the Weyl anomaly we obtained in \eqref{8dA} when the Weyl structure is turned off, up to total derivatives. 
\end{appendix}

\begingroup\raggedright
\endgroup
\end{document}